\documentclass{ws-ijmpe}
\usepackage[super,compress]{cite}
\usepackage{color}
\usepackage[utf8]{inputenc}
\usepackage[english]{babel}
\usepackage{hyperref}
\usepackage[figuresright]{rotating}
\begin{document}
\markboth{R. Sharma \textit{et al.}}{Structural properties of nuclei with semi-magic number N(Z)$=$40}
\catchline{}{}{}{}{}
%%%%%%%%%%%%%%%%%%%%%%%%%%%%%%%%%%%%%%%%%%%%%%%%%%%%%%%%%%%%%%%%%%%%
\title{Structural properties of nuclei with semi-magic number N(Z)$=$40}
\author{R. Sharma}
\address{{Department of Physics, School of Basic Sciences, Manipal University Jaipur, Jaipur-303007, India},\\
  {Department of Physics, S. S. Jain Subodh P.G.(Autonomous) College, Jaipur-302004, India}}
\author{A. Jain}
\address{{Department of Physics, School of Basic Sciences, Manipal University Jaipur, Jaipur-303007, India},\\
  {Department of Physics (H $\&$ S), Govt. Women Engineering College, Ajmer-305002, India}}
\author{M. Kaushik}
\address{S. S. Jain Subodh P. G. College, M. C. A. Institute, Rambagh Circle, Jaipur-302004, India}
\author{S. K. Jain}
\address{Department of Physics, School of Basic Sciences, Manipal University Jaipur, Jaipur-303007, India}
\author{G. Saxena}
\address{Department of Physics (H $\&$ S), Govt. Women Engineering College, Ajmer-305002, India\\ gauravphy@gmail.com}

\maketitle

\begin{history}
\received{Day Month Year}
\revised{Day Month Year}
%\accepted{Day Month Year}
%\comby{(xxxxxxxxxx)}
\end{history}
\begin{abstract}
Various ground state properties are explored for full isotonic(isotopic) chain of neutron number N(proton number Z)$=$40 using different families of Relativistic Mean-Field theory. Several properties such as nucleon separation energies, pairing energies, deformation, radii and nucleon density distributions are evaluated and compared with the experimental data as well as those from other microscopic and macroscopic models. N$=$40 isotonic chain presents ample of support for the neutron magicity and articulates double magicity in recently discovered $^{60}$Ca and $^{68}$Ni. Our results are in close conformity with recently measured value of charge radius of $^{68}$Ni [S. Kaufmann \textit{et al.}, Phys. Rev. Lett. 124, 132502 (2020)] which supports the N$=$40 magicity. Contrarily, Zr isotopes (Z$=$40) display variety of shapes leading to the phenomenon of shape transitions and shape co-existence. The role of 3s$_{1/2}$ state, which leads to central depletion if unoccupied, is also investigated. $^{56}_{16}$S and $^{122}_{40}$Zr are found to be doubly bubble nuclei.
\end{abstract}

%%%%%%%%%%%%%%%%%%%%%%%%%%%%%%%%%%%%%%%%%%%%%%%%%%%%%%%%%%%%%%%%%%%%%%%%%%%%%%%%%%%%%%%%%%%%%%%%%%%%%%%%%%%%%%%%%%%%%%%%%%%%%%%%%%%%%%%%%%%%%%%%%%%%%%%
\section{Introduction}
The recent discovery of $^{60}_{20}$Ca$_{40}$ at the radioactive ion-beam factory
operated by RIKEN Nishina Center and CNS, University of Tokyo \cite{tarasov2018} and recent lifetime measurements demonstrating shape-coexistence in $^{98}_{40}$Zr$_{58}$ \cite{purnima2018} have certainly provided an extra impulse to ascent more theoretical and experimental studies for the nuclei consisting neutron(proton) number N(Z)$=$40. Latterly, the first spectroscopy of $^{62}$Ti has been reported for the detail investigation of shell evolution of N$=$40 isotones towards $^{60}$Ca \cite{cortes2020}. Another nucleus $^{68}_{28}$Ni$_{40}$ has shown tremendous possibilities to look into the magicity \cite{izzo2018}, shapes and shape-coexisting configurations in excited states \cite{flavigny2019}, giant resonances \cite{bonasera2018} and pygmy dipole resonance (PDR) \cite{martorana2018,sun2018}. Evolution of structure in the Zr isotopes (Z$=$40) has been described to show the interplay between shape-phase transitions and shape coexistence \cite{gavrielov2019}.

Full systematic study using covariant density functional analysis of N$=$40 isotones has been done by Wang \textit{et al.} \cite{wang2015} which provides (a) reasonable description not only for the systematics of the low-lying states along the isotonic chain but also for the detailed structure of the spectroscopy in a single nucleus (b) spherical-oblate-prolate shape transition along the isotonic chain of N$=$40 and, (c) coexistence of low-lying excited ${{0}^{+}}$ states in neutron-deficient N$=$40 isotones. On the other hand, shape coexistence in Zr isotopes has been investigated using interacting boson model with configuration mixing \cite{ramos2019} which has confirmed spherical nature of the ground state of $^{94-98}$Zr and deformed nature for $^{100-110}$Zr. Moreover, role of tensor force is inquired for the shape evolution of Zr nuclei using axially deformed Hartree-Fock (HF) calculations with the semi-realistic interaction M3Y-P6 \cite{miyahara2018} and as a result a strong shape transition is found in Zr isotopes such as $^{80}$Zr is found deformed whereas $^{86-96}$Zr are found spherical. Prolate shape of $^{98-112}$Zr switches to oblate for $^{114}$Zr and then sphericity returns at $^{120}$Zr and $^{122}$Zr \cite{miyahara2018}. Therefore, Zr isotopes exhibit remarkable N-dependence in their structure leading to a demand of full systematic and deeper study which is one of the prime reason for this investigation. The existence of a new island of inversion around N$=$40 \cite{adrich2008,ljungvall2010,pauwels2008,lenzi2010,gaude2009,naimi2012} along with recent experimental studies \cite{liu2018,wimmer2019}, waiting point nuclei along N$=$Z line for rapid proton capture reactions (rp-process) \cite{brown2002}, evolution of N(Z)$=$40 \cite{otsuka2020} along with new doubly magic nuclei $^{60}$Ca \cite{tarasov2018} and $^{68}$Ni \cite{izzo2018} are the driving forces which have prompted this systematic study of nuclei with (semi) magic number N(Z)$=$40.\par
We perform a systematic study of full chain of N$=$40 isotones and Z$=$40 (Zr) isotopes using relativistic mean-field (RMF) models. Majority of results are presented using the RMF model with density dependent meson coupling strength taking DD-ME2 parameter \cite{Lalazissis05}. At various places these results are compared with RMF models with density-dependent point coupling interactions using
DD-PCX parameter \cite{ddpcx} and the nonlinear self- and mixed-interactions of the mesons viz. NL3* and FSU-Gold \cite{Boguta77,Boguta83,Furnstahl97,Todd-Rutel05,Lalazissis09} and experimental data as well \cite{nndc,angeli}. This systematic study consist of ground state properties which include deformation, separation energies, single particle energies, radii and density distribution etc.

\section{Relativistic Mean-Field Theory}
Three classes of mean-field models are employed for this investigation: the density-dependent meson-exchange (DD-ME) model, the nonlinear meson-nucleon coupling model (NL), and the density-dependent point-coupling (DD-PC) model. The main differences between these models consist in the treatment of the range of the interaction and in the density dependence. The density dependence is introduced either through an explicit dependence of the coupling constants (DD-ME \cite{Lalazissis05} \& DD-PC \cite{ddpcx}) or via non-linear meson couplings (NL \cite{Lalazissis09}). Meson exchange model and point coupling model have an interaction of finite and of zero range, respectively. As a consequence, at present, these major classes of covariant energy density functionals exist dependent on the combination of above mentioned features (see Ref. \cite{agbemava2014} for detail).\par

\subsection{Variants of RMF}
We perform the RMF calculations using effective nuclear interactions with density-dependent meson-nucleon vertex functions which represent a significant improvement in the relativistic self-consistent mean-field description of the nuclear many-body problem. This kind of effective interaction has been shown to provide a more realistic description of asymmetric nuclear matter, neutron matter, and finite nuclei, which includes a softer equation of state of nuclear matter (i.e., lower incompressibility) and a lower value of the symmetry energy at saturation in comparison to other standard nonlinear meson-exchange models \cite{Lalazissis05}. The employed effective Lagrangian density is similar to the Ref.$~$ \cite{Lalazissis05} and has the following form:
\begin{eqnarray}
{\cal L} &=& \bar{\psi}(\imath\psi.\partial - M)\psi + \frac{1}{2}  (\partial\sigma)^2 - \frac{1}{2}m^2_\sigma \sigma^2 - \frac{1}{4}\bf\Omega_{\mu\nu}\bf\Omega^{\mu\nu}\nonumber\\
 &&+ \frac{1}{2}m^2_\omega\omega^2
-\frac{1}{4}\overrightarrow{\bf{R}}_{\mu\nu}\overrightarrow{\bf{R}}^{\mu\nu}+\frac{1}{2}m_\rho^2\overrightarrow{\rho}^2-\frac{1}{4}\bf{F}_{\mu\nu}\bf{F}^{\mu\nu}\nonumber \\
&& -g_\sigma\bar{\psi}\sigma\psi - g_\omega\bar{\psi}\gamma .\omega\psi-g_\rho\bar{\psi}\gamma .\overrightarrow{\rho}\overrightarrow{\tau}\psi\nonumber\\
&&-\epsilon\bar{\psi}\gamma . {\bf{A}} \frac{(1-\tau_3)}{2}\psi
\end{eqnarray}
Here, vectors in isospin space are denoted by arrows, and the bold-faced symbols indicate vectors in ordinary three-dimensional space. The Dirac spinor $\psi$ denotes the nucleon with mass M. The masses of $\sigma$, $\omega$, and the $\rho$ mesons are denoted by m$_{\sigma}$,  m$_{\omega}$, and m$_{\rho}$, respectively, with g$_{\sigma}$, g$_{\omega}$, and g$_{\rho}$ being the corresponding coupling constants for the mesons to the nucleon. Here, $\bf\Omega^{\mu\nu}$,  $\overrightarrow{\bf{R}}^{\mu\nu}$, and $\bf{F}^{\mu\nu}$ are the field tensors of the vector fields $\omega$, $\rho$, and of the photon:
\begin{eqnarray}
                \bf\Omega^{\mu \nu} &=& \partial^{\mu} \omega^{\nu} -
                      \partial^{\nu} \omega^{\mu}\nonumber\\
                \overrightarrow{\bf{R}}^{\mu\nu} &=& \partial^{\mu} \overrightarrow{\rho}^{\nu} -
                       \partial^{\nu} \overrightarrow{\rho}^{\mu}\nonumber\\
                \bf{F}^{\mu \nu} &=& \partial^{\mu} \bf{A}^{\nu} -
                      \partial^{\nu} \bf{A}^{\mu}
\end{eqnarray}

As per Ref.$~$\cite{Lalazissis05} the density-dependent meson-exchange model (DD-ME)
is used in this work, in which the meson-nucleon strengths $g_{\sigma}$, $g_{\omega}$ and $g_{\rho}$ have an explicit density
dependence in the following form:

\begin{equation} \label{eq:Equation1} %\begin{split}
g_{i}(\rho) =  g_{i}(\rho_{sat})f_{i}(x), \,\,\,\,\,\,\, for \,\,\, i = \sigma, \omega %\end{split}
\end{equation}

where the density dependence is given by:
\begin{equation} \label{eq:Equation2} %\begin{split}
f_{i}(x) =  a_{i} \frac {1+b_{i}(x+d_{i})^{2}}{1+c_{i}(x+d_{i})^{2}} %\end{split}
\end{equation}
in which $x$ is given by $x = \rho/\rho_{sat}$, and $\rho_{sat}$ denotes the baryon density
at saturation in symmetric nuclear matter.
For the $\rho$ meson, density dependence is of exponential form and given by
\begin{equation} \label{eq:Equation3} %\begin{split}
f_{\rho}(x) =  exp(-a_{\rho}(x-1)) %\end{split}
\end{equation}

For a comparison, we have used an improved parameterization of widely used and successful non-linear parameter (NL3$^{\ast}$) for the RMF model, which contains only six phenomenological parameters. This variant has delivered accurate description of the ground state properties of many nuclei and simultaneously has provided an excellent description of excited states with collective character in spherical as well as in deformed nuclei \cite{Lalazissis09}. This variant includes linear terms for the $\sigma$, $\omega$ and $\rho$ mesons, together with the non-linear term only for self interaction of the $\sigma$ meson. The effective Lagrangian density can be expressed by adding the nonlinear term of $\sigma$ meson ($\frac{1}{3}g_{2}\sigma^{3} + \frac{1}{4}g_{3}\sigma^{4}$). For more details see Refs.$~$\cite{Boguta77,Boguta83,Furnstahl97,Lalazissis09}.\par

Our calculations are also performed with the nonlinear parameter FSU-Gold \cite{Todd-Rutel05} containing non-linear self interaction of $\omega$ meson and the mixed interaction terms for $\omega$ and $\rho$ mesons. These two additional parameters are used by virtue of softening of both the EOS of symmetric matter and the symmetry energy. This variant FSU-Gold has been fitted to the binding energies and charge radii of a variety of magic nuclei. The interaction part of the Lagrangian, which describes the coupling of mesons to the nucleons and the non-linear self and mixed interactions of mesons, can be expressed as \cite{Todd-Rutel05}:
\begin{eqnarray} \label{eq:Lagrangian2}
 %\begin{split}
{\cal L}_{\it int}=&&\overline{\psi}\left [g_{\sigma} \sigma -\gamma^{\mu} \left (g_{\omega }
\omega_{\mu}+\frac{1}{2}g_{\mathbf{\rho}}\tau .
\mathbf{\rho}_{\mu}+\frac{e}{2}(1+\tau_3)A_{\mu}\right ) \right ]\psi\nonumber
\\ &&-\frac{{\kappa_3}}{6M}
g_{\sigma}m_{\sigma}^2\sigma^3-\frac{{\kappa_4}}{24M^2}g_{\sigma}^2
m_{\sigma}^2\sigma^4+ \frac{1}{24}\zeta_0
g_{\omega}^{2}(\omega_{\mu}\omega^{\mu})^{2}\nonumber
\\ &&+\frac{\eta_{2\rho}}{4M^2}g_{\omega}^2m_{\rho
}^{2}\omega_{\mu}\omega^{\mu}\rho_{\nu}\rho^{\nu}.
%\end{split}
\end{eqnarray}

It includes an isoscalar-scalar $\sigma$ meson field and three vector fields: an isoscalar $\bf{\omega}_\mu$, an isovector $\bf{\rho}_\mu$, and the photon $\bf{A}_\mu$. In addition to the Yukawa couplings, the Lagrangian is supplemented by four nonlinear meson interactions. The inclusion of isoscalar meson self-interactions (via $\kappa_3$, $\kappa_4$, and $\zeta_0$) are used to soften the equation of state of symmetric nuclear matter, while the mixed isoscalar-isovector coupling ($\eta_{2p}$) modifies the density dependence of the symmetry energy. Other symbols have the usual meaning and details can be found in Refs.$~$ \cite{Boguta77,Boguta83,Furnstahl97,Todd-Rutel05}.\par

In analogy with meson-exchange model (DD-ME) described above, we have used density-dependent point coupling interaction (DD-PC) \cite{ddpcx} in the RMF calculations.
The effective interaction DD-PCX used in this work represents the first effective interaction that is constrained using the binding energies, charge radii,
and pairing gaps, together with a direct implementation of the ISGMR energy and dipole polarizability \cite{ddpcx}. This variant accurately describes the nuclear ground state properties including the neutron-skin thickness, as well as the isoscalar giant monopole resonance excitation energies and dipole polarizabilities. The effective lagrangian for this model in terms of nucleonic field can be expressed as:
\begin{eqnarray} \label{eq:Lagrangian3} %\begin{split}
{\cal L}=&& \overline{\psi}(\imath\gamma . \partial - M)\psi\nonumber\\
&&-\frac{1}{2}\alpha_{S}(\rho)(\overline{\psi}\psi)(\overline{\psi}\psi) -
\frac{1}{2}\alpha_{V}(\rho)(\overline{\psi}\gamma^{\mu}\psi)(\overline{\psi}\gamma_{\mu}\psi)\nonumber
\\ &&- \frac{1}{2}\alpha_{TV}(\rho)(\overline{\psi}\overrightarrow{\tau}\gamma^{\mu}\psi)(\overline{\psi}\overrightarrow{\tau}\gamma_{\mu}\psi) - \frac{1}{2}\delta_{S}(\partial_{\nu}\overline{\psi}\psi)(\partial^{\nu}\overline{\psi}\psi)\nonumber\\
&& - \epsilon\overline{\psi}\gamma . \bf{A}\frac{(1-\tau_3)}{2}\psi %\end{split}
\end{eqnarray}

Here, the free nucleonic term contains isoscalar-scalar (S), isoscalar-vector (V) and isovector-vector (TV) interactions. The coupling constants $\alpha_{i}(\rho)$ are density dependent and employed as:
\begin{equation} \label{eq:Equation4} %\begin{split}
\alpha_{i}(\rho)  =  a_{i} + (b_{i} + c_{i}x)e^{-d_{i}x},    for \,\,\, i = S, V, TV%\end{split}
\end{equation}
where $x = \rho/\rho_{o}$, and $\rho_{o}$ denotes the nucleon density in symmetric nuclear matter at saturation point. It should be pointed out here that success of the DD-PCX interaction in the predictions of the dipole polarizabilities and neutron skin thicknesses in other nuclei not used in optimizing the model parameters validates the isovector channel of the functional and the respective symmetry energy properties.

\subsection{Pairing Treatment}
Pairing correlations play an important role in all open-shell nuclei and turn into particulary significant for the nuclei near the drip line. For the mean-field level pairing correlations are taken into account by Bardeen-Cooper-Schrieffer (BCS) or HartreeFock-Bogoliubov (HFB) theory and in the relativistic case by relativistic Hartree-Bogoliubov (RHB) theory \cite{kucharek1991,ring1996,afanasjev2000}. For the non-linear version of our calculations (NL3*), we have used the BCS scheme wherein the single particle continuum corresponding to the RMF is replaced by a set of discrete positive energy states. The results are found to be in close
agreement with the experimental data and with those of recent continuum relativistic Hartree-Bogoliubov (RCHB) and other similar mean-field calculations \cite{saxena4}. In the calculations, we use a delta force for the pairing interaction, i.e., $V = -V_{0} \delta(r)$ with the strength $V_{0} = 350$ MeV fm$^3$ which has been used in Refs. \cite{Yadav2004,saxena4} for the successful description of drip-line nuclei. Apart from its simplicity, the applicability and justification of using such a $\delta$-function form of interaction has been discussed in Ref. \cite{Dobaczewski1983}, whereby it has been shown in the context of HFB calculations that the use of a delta force in a finite space simulates the effect of finite range interaction in a phenomenological manner (see also \cite{Bertsch1991} for more details). \par

Whenever the zero-range $\delta$ force is used either in the BCS or the Bogoliubov framework, a cutoff procedure must be applied, i.e. the space of the single-particle states where the pairing interaction is active must be truncated. This is not only to simplify the numerical calculation but also to simulate the finite-range (more precisely, long-range) nature of the pairing interaction in a phenomenological way \cite{Dobaczewski1995}. In the present work, the single-particle states subject to the pairing interaction are confined to the region satisfying
\begin{equation}
\epsilon_i-\lambda\le E_\mathrm{cut},
 \end{equation}
where $\epsilon_i$ is the single-particle energy, $\lambda$ the Fermi energy, and $E_\mathrm{cut}$=$8.0$ MeV.
%The center-of-mass correction is approximated by
% \begin{equation}
% E_{\textrm{cm}} = -\frac{3}{4}41A^{-1/3},
% \end{equation}
%which is often used in the relativistic mean-field theory among the many recipes for the center-of-mass correction
%\cite{Bender1999}.
For further details of these formulations we refer the
reader to Refs. \cite{Gambhir1989,Singh2013,Geng2003}. \par

The pairing correlations for FSU-Gold lagrangian are treated in the constant gap approximation \cite{reinhard1986,suga1994} with gap parameters derived from the odd-even mass differences. This approach works properly when the main effect of the pairing correlations is a smearing of the Fermi surface. The constant pairing gaps are taken as the following forms
 \begin{equation}
 \triangle_n = \triangle_p = \frac{11.2}{\sqrt{A}}
  \end{equation}

For the density dependent models, we use the TMR separable pairing force of Ref. \cite{TMR} for the short range correlations. This kind of separable pairing force has been adjusted to reproduce the pairing gap of the Gogny force D1S in symmetric nuclear matter. Both forces are of finite range and therefore they show no ultraviolet divergence and do not depend on a pairing cut-off. They provide a very reasonable description of pairing correlations all over the periodic table with a fixed set of parameters. In the $^{1}$S$_{0}$ channel the gap equation is given by
 \begin{equation}
\Delta(k) = \int_{0}^{\infty} \frac{k'^2 dk'}{2\pi^2} \langle k|V^{^{1}S_{0}}|k'\rangle \frac{\Delta k'}{2E(k')}
  \end{equation}
  and the pairing force separable in momentum space is
 \begin{equation}
  \langle k|V^{^{1}S_{0}}|k'\rangle = - Gp(k)p(k')
  \end{equation}
The two parameters determining the force are, the pairing strength G and $\alpha$ that goes in the Gaussian ansatz $p(k)= e^{-\alpha^2 k^2}$. Their value has been adjusted to G = 728 MeV fm$^3$ and $\alpha$ = 0.644 fm in order to reproduce the density dependence of the gap at the Fermi surface, calculated with the D1S parametrization of the Gogny force \cite{berger1991}.

%-------------------------------------------------------------------------------------------------
\section{Results and discussions}
Study of N$=$40 isotones covers a chain of proton number from Z$=$16 to Z$=$42 whereas Zr isotopes (Z$=$40) contain a wide range of neutrons from N$=$38 to N$=$112 which in turn address to a broad area of periodic chart. We first describe shapes and structural properties of the nuclei of these chains which are followed by a detailed investigation of observed magicity. Thereafter, central depletion or bubble phenomenon in the nucleonic density is examined.

\subsection{Shapes and Structural Properties}
\begin{figure}[h]
\centering
\includegraphics[width=0.9\textwidth]{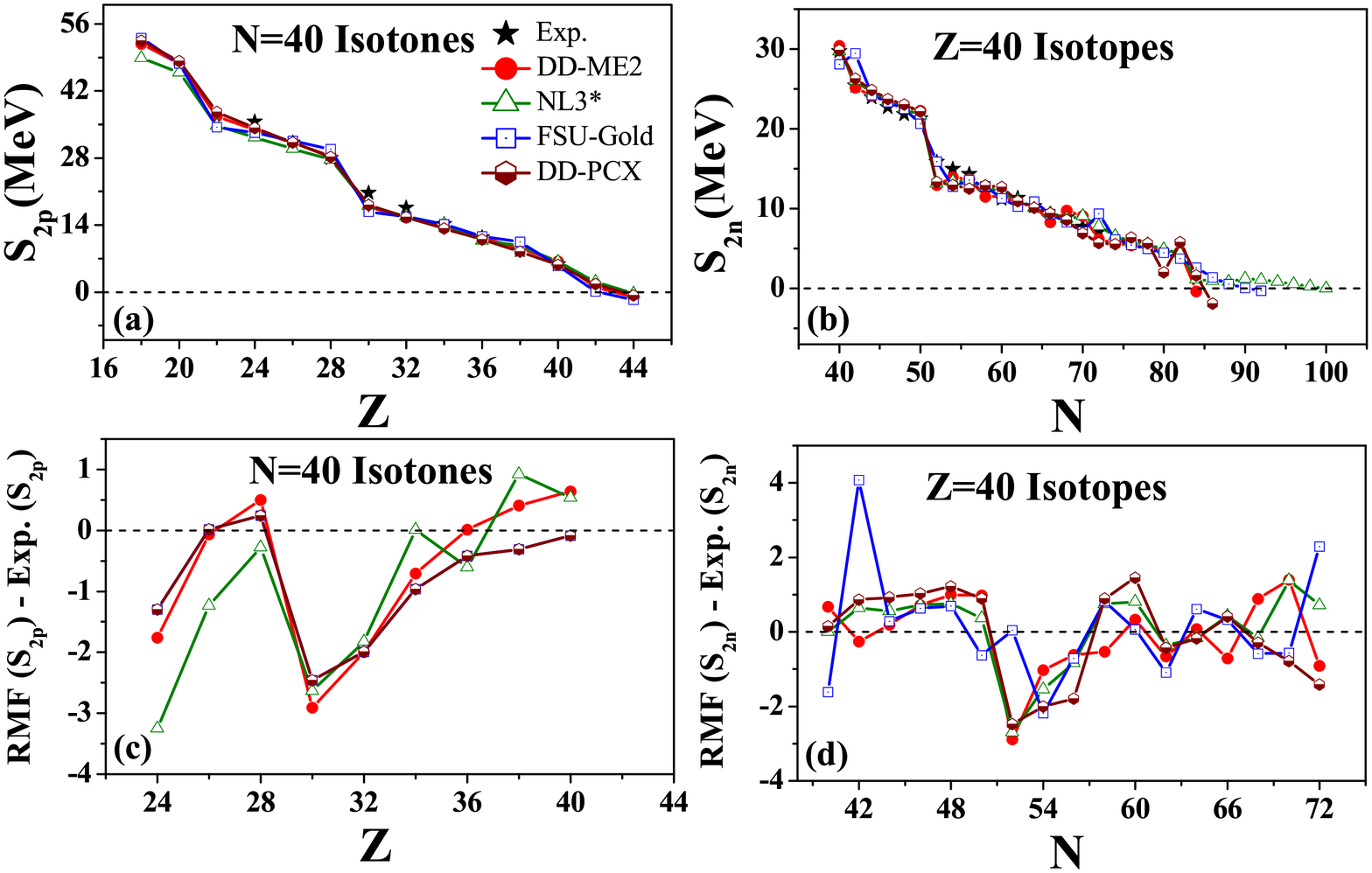}
\caption{(Colour online) Two proton separation energies (S$_{2p}$) and two neutron separation energies (S$_{2n}$) calculated with DD-ME2, NL3*, FSU-Gold, and DD-PCX, parameters along with experimental separation energies \cite{nndc} for N$=$40 isotones and Z$=$40 isotopes are shown in panels (a) and (b), respectively. Difference between theoretical results with experimental data are also shown in panels (c) and (d). A reasonable agreement of theoretical results with experimental data is clearly visible in panels (a) and (b). A sharp fall in the values of S$_{2p}$ and S$_{2n}$ just beyond the magic number can also be looked at. Panels (c) and (d) results a closer agreement of DD-ME2 parameter as a whole.} \label{fig1}
\end{figure}

We have plotted two proton separation energy (S$_{2p}$) of N$=$40 isotones and two neutron separation energy (S$_{2n}$) of Zr isotopes in Fig. \ref{fig1} (a) and (b), respectively to validate our results. These energies are calculated using DD-ME2 parameter~\cite{Lalazissis05} and compared also with the RMF calculations done with NL3* \cite{Lalazissis09}, FSU-Gold \cite{Todd-Rutel05} and DD-PCX \cite{Lalazissis09} along with experimental data \cite{nndc}. A reasonable match among all these parameters along with experimental results can be seen from Fig. \ref{fig1}(a) for N$=$40 isotones. Two proton drip-line for this chain is found $^{82}$Mo from DD-ME2 and DD-PCX parameter whereas NL3* and FSU-Gold  parameters deliver to $^{80}$Zr. A sharp fall just after conventional magic numbers Z$=$20 and 28 is clearly observed affirming the magicity in $^{60}$Ca and $^{68}$Ni: the nuclei which are of current interest \cite{tarasov2018,izzo2018} as mentioned above.\par

From Fig. \ref{fig1}(b), for Zr isotopes, a clear-cut dependency on the chosen parameter can be seen. To visualize this variance, we have plotted the differences of our theoretical estimates with available experimental values of S$_{2p}$ and S$_{2n}$ for N$=$40 isotones and Zr isotopes in Fig. \ref{fig1} (c) and (d), respectively. By and large, the results of DD-ME2 parameter are found in closest match with experimental data as compared to the other parameters. From density dependent parameters two-neutron drip line for Zr isotopes is reported as $^{122}$Zr, however, from the non-linear parameters i.e NL3* and FSU-Gold, it is found to extend upto a larger distance at $^{138}$Zr and $^{130}$Zr, respectively. It is tempting to associate the difference in the position of two-neutron drip lines with different symmetry energies J (J = 32.30, 38.6, 32.6 and 31.12 MeV for DD-ME2, NL3*, FSU-Gold and DD-PCX, respectively) and the slope parameter L of the symmetry energy at saturation density (L = 51.26, 122.6, 60.5, 46.32 MeV for DD-ME2, NL3*, FSU-Gold and DD-PCX, respectively). A comparison of these nuclear matter properties does not clearly reveal any correlations with the locations of two-neutron drip lines as also has been described in Ref \cite{agbemava2014}. In fact, the precise position of the drip line depends very much on the behavior of the tail of the neutron density which is not really relevant in comparison to the considered values of J and L of nuclear matter at saturation. However, the difference in the position of drip-line may be attributed to the pairing treatment adopted for the calculations. For an instance, the extension in neutron drip-line using state dependent BCS pairing has been explained by Yadav \textit{et al.} \cite{Yadav2004} and Meng \textit{et al.} \cite{meng1998} with the comment that for the drip-line nuclei the role of continuum states and their coupling to the bound states become exceedingly important, especially for the pairing energy contribution to the total binding energy of the system. The scattering of particles from bound to continuum states near the Fermi level and vice versa due to the pairing interaction pushes the drip-line towards more neutron side with a small increase in binding energy. Contrary to the variance in drip-line, sharp fall just after neutron number N$=$50 and 82 from all the considered parameters demonstrate strong magicity for $^{90}$Zr and $^{122}$Zr, respectively in accord with recent studies \cite{gavrielov2019,miyahara2018}.

\begin{figure*}[h]
\centering
\includegraphics[width=0.7\textwidth]{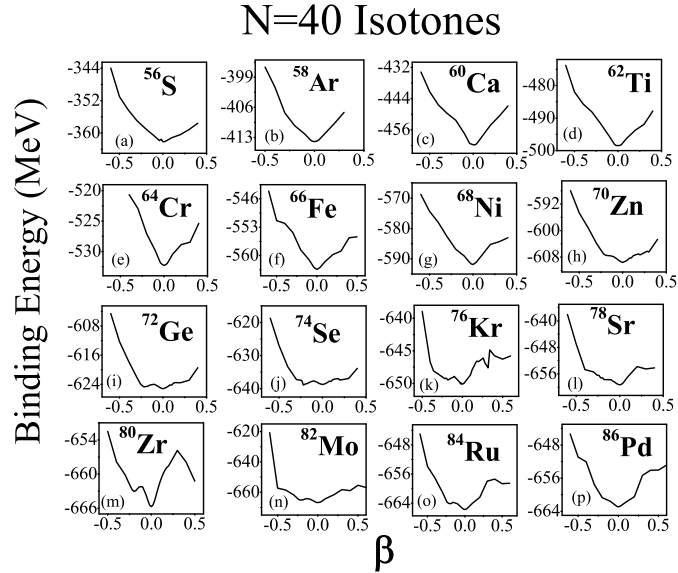}
\caption{The potential energy surfaces of N$=$40 isotones as a function of the deformation parameter $\beta$ calculated by DD-ME2 parameter. Spherical shapes of $^{60}$Ca and $^{68}$Ni suggest magic character whereas few other nuclei are found with very flat minima or with a secondary minima which consequently lead to chances of shape co-existence.} \label{fig2}
 \end{figure*}
\begin{figure*}[h]
\centering
\includegraphics[width=0.95\textwidth]{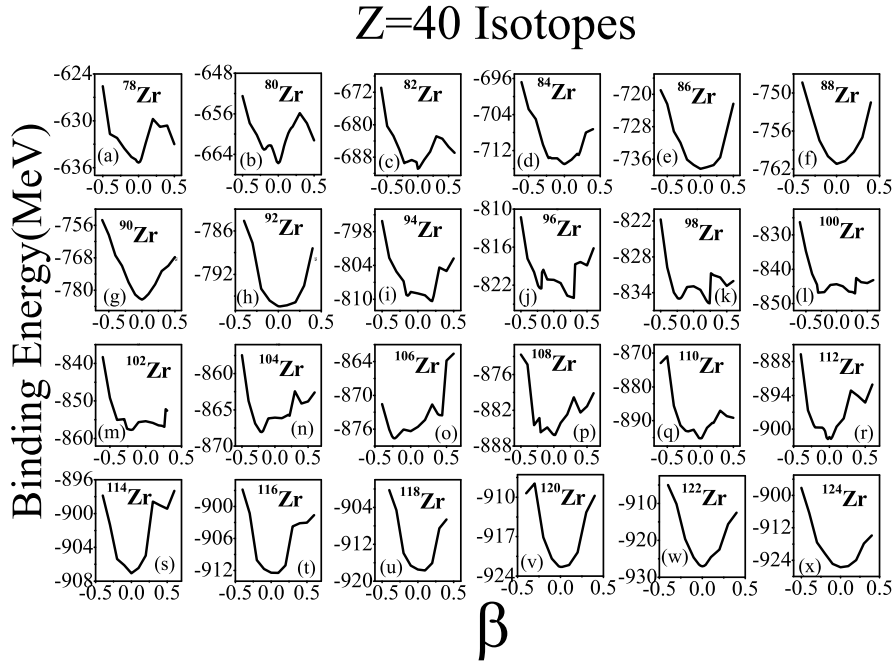}
\caption{The potential energy surfaces of Zr (Z$=$40) isotopes as a function of the deformation parameter $\beta$ calculated by DD-ME2 parameter. $^{90}$Zr and $^{122}$Zr are found with a clean and single minima at zero deformation which establishes their spherical shapes and magicity. Many other isotopes are found with deformed shapes or with the coexisting shapes.} \label{fig3}
\end{figure*}

To look into the shapes of these nuclei, in Figs. \ref{fig2} and \ref{fig3}, potential energy surfaces calculated by DD-ME2 parameter are shown as a function of deformation $\beta$ for N$=$40 isotones and Zr isotopes, respectively. Many of the potential energy surfaces of N$=$40 isotones as can be seen from Fig. \ref{fig2} result to dominant ground state configurations with spherical shapes ensuing magic character of N$=$40, out of which $^{60}$Ca and $^{68}$Ni appear once again with eminent sphericity. However, a close examination of Fig. \ref{fig2} shows that after $^{68}$Ni, for the nuclei between $^{70}$Zn and $^{84}$Ru the ground state configurations with only spherical shape are not so dominant as found for the cases before $^{68}$Ni upto $^{56}$S. The minima are either found very flat or with a few secondary minima which consequently lead to chances of shape co-existence. These nuclei are mentioned in Table \ref{beta} with the excitation energy: the difference of energy between two minima along with possible shape due to secondary minima (O) for oblate secondary minima and (P) for prolate secondary minima. It is gratifying to note that nuclei  $^{72}$Ge, $^{74}$Se, $^{76}$Kr, $^{78}$Sr, $^{80}$Zr, $^{82}$Mo, $^{84}$Ru are indeed found as the possible candidates of shape co-existence in accord with various Refs.~\cite{nomura,ayangeakaa,akbari}. It is important to point out here that for the case of $^{80}$Zr, spherical minima is found more dominant similar to what has been found in various other mean-field calculations \cite{miyahara2018,delaroche2010}, whereas $^{80}$Zr is indeed deformed as has been described in various Refs. \cite{akbari,Bender09,Zou2010,Zheng2014}. Nevertheless, the FSU-Gold parameter renders fairly large deformation for the case of $^{78,80}$Zr which may be attributed to the different choice of pairing in this variant of RMF. For other parameters, we have recalculated potential energy surface for $^{80}$Zr by increasing pairing strength and it has been found that the spherical minima goes flatter and the deformed minima becomes more dominant, which indicates need of deeper study along this direction. With this example of $^{80}$Zr near drip-line together with the variance in exact position of drip-line, it is concluded that for drip-line nuclei the treatment of pairing becomes very crucial which in turn leaves at somewhat distinct results from different variants of theoretical approaches.\par

In a similar manner, from Fig. \ref{fig3} one can explore the shapes for full chain of Zr isotopes. As expected, $^{90}$Zr and $^{122}$Zr are found with a clean and single minima at zero deformation registering their spherical shapes similar to what observed by Hartree-Fock (HF) calculations with the semi-realistic interaction M3Y-P6 \cite{miyahara2018}. On the other hand, $^{86,88,92}$Zr and $^{112-120}$Zr are found with either flatter minima or mainly with one shape. In connection to the recent measurements \cite{purnima2018,gavrielov2019}, from our theory also many of Zr isotopes are found to show shape co-existence. These nuclei $^{82,84,94,96,98,100,102,106,108}$Zr are mentioned in Table \ref{beta} with their excitation energy and are found in agreement with Refs.~\cite{purnima2018,miyahara2018,akbari,Sazo,witt,kremer,YanXin} to register their candidature as nuclei showing shape co-existence. As mentioned above, in Table \ref{beta}, (O) is used for oblate secondary minima and (P) for prolate secondary minima if dominant minima is spherical, whereas (O') and (P') refer to the oblate secondary and for prolate secondary minima while dominant minima is prolate and oblate, respectively.

\begin{table*}[!htbp]
\caption{Results of excitation energy (energy difference between two minima) as obtained in the RMF calculations using DD-ME2 force parameter for N(Z)$=$40 Isotones(Isotopes). The table clearly demonstrates many number of nuclei which are potential candidates showing shape-coexistence.}
\centering
\resizebox{0.55\textwidth}{!}{%
\begin{tabular}{|c|c|c|}
\hline Nucleus &Excitation Energy (MeV)&Other references\\
\hline
$^{72}${Ge}&0.55 (O)&\cite{nomura,ayangeakaa}\\
$^{74}${Se}&0.17 (O)&\cite{nomura}\\
$^{76}${Kr}&0.65 (O), 2.51 (P)&\cite{akbari}\\
$^{78}${Sr}&0.45 (O), 0.54 (P)&\cite{akbari}\\
$^{80}${Zr}&2.67 (O)&\cite{akbari}\\
$^{82}${Mo}&1.64 (O)&\\
$^{84}${Ru}&1.60 (O)&\\
\hline
$^{82}${Zr}&1.38 (O)&\cite{akbari}\\
$^{84}${Zr}&1.34 (O)&\\
$^{94}${Zr}&0.95 (O') &\\
$^{96}${Zr}&1.49 (O')&\cite{Sazo,witt,kremer}\\
$^{98}${Zr}&0.80 (O')&\cite{purnima2018,witt}\\
$^{100}${Zr}&0.21 (O')&\cite{witt}\\
$^{102}${Zr}&0.85 (P')&\cite{miyahara2018}\\
$^{106}${Zr}&1.28 (P')&\\
$^{108}${Zr}&0.45 (O)&\cite{YanXin}\\
\hline
\end{tabular}}
\label{beta}
\end{table*}

\begin{table*}
\caption{Ground state quadrupole deformation $\beta$ calculated with DD-ME2 parameter for the nuclei with N(Z)$=$40 are mentioned and compared with other considered parameters NL3*, FSU-Gold, and DD-PCX. These deformations are also compared with other theoretical model/parameter viz. RMF(TMA) \cite{geng2004}, FRDM \cite{frdm2012}, Skyrme-Hartree-Fock-Bogoliubov mass formulas (HFB) using HFB-24 \cite{goriely,goriely-xu}, Hartree-Fock-Bogoliubov approach using the Gogny D1S effective interaction (D1S for N$=$40 isotones) \cite{gaude2009}, and Skyrme-Hartree-Fock-Bogoliubov formalism with SLy4 parameter (SLy4 for Zr isotopes) \cite{bharat2015}.}
\centering
\resizebox{0.9\textwidth}{!}{%
{\begin{tabular}{|c|c|c|c|c|c|c|c|c|c|}
\hline
\multicolumn{09}{|c|}{N$=$40 Isotones}\\
\hline
\multicolumn{1}{|c}{Nucleus}&
\multicolumn{1}{|c}{DD-ME2}&
\multicolumn{1}{|c}{NL3$^*$}&
 \multicolumn{1}{|c}{FSU}&
 \multicolumn{1}{|c}{DD-PCX}&
  \multicolumn{1}{|c}{TMA}&
  \multicolumn{1}{|c}{FRDM}&
  \multicolumn{1}{|c}{D1S}&
  \multicolumn{1}{|c|}{HFB}\\
  %\multicolumn{1}{|c|}{Expt.}\\
 \hline
 $^{56}${S} &0.00 &-0.02&0.05&0.00    &0.00&-0.22 &&            \\
 $^{58}${Ar}&0.00 &0.00 &0.00&0.00  & 0.00&-0.22 &0.00&0.09          \\
 $^{60}${Ca}&0.00 &0.00 &0.00&0.00  & 0.00&0.00  &0.00&0.00        \\
 $^{62}${Ti}&0.00 &0.00 &0.04&0.00  & 0.00&0.00  &0.00&0.11      \\
 $^{64}${Cr}&0.00 &0.00 &0.06&0.00  & 0.00&0.00  &0.00&0.19   \\
 $^{66}${Fe}&0.00 &0.00 &0.05&0.00  & 0.00&0.00  &0.00&0.10   \\
 $^{68}${Ni}&0.00 &0.00 &0.00&0.00  & 0.00&0.00  &0.00&0.09   \\
 $^{70}${Zn}&0.00 &0.00 &0.05&0.00  & 0.00&0.00  &0.00&-0.16 \\
 $^{72}${Ge}&0.00 &-0.20&0.06&0.00    &-0.21&-0.22&-0.2&-0.24 \\
 $^{74}${Se}&0.00 &-0.24&0.00&0.00    &-0.23&-0.24&-0.2&-0.23\\
 $^{76}${Kr}&0.00 &0.00 &0.04&0.00   & -0.32&0.40 &0.00&-0.24\\
 $^{78}${Sr}&0.00 &0.49 &0.00&0.00  & -0.49&0.40 &0.00&-0.22 \\
 $^{80}${Zr}&0.00 &0.00 &0.46&0.00  & 0.49 &0.43 &0.00&-0.23    \\
 $^{82}${Mo}&0.00 &0.00 &0.03&0.00  & 0.59 &0.47 &0.00&-0.22    \\
 $^{84}${Ru}&0.00 &0.00 &0.06&0.00  & -0.22&-0.23&&0.06              \\
 $^{86}${Pd}&0.00 &0.00 &0.07&0.00  & -0.17&-0.24&&-0.04            \\
 \hline
 \multicolumn{09}{|c|}{Z$=$40(Zr) Isotones}\\
 \hline
 \multicolumn{1}{|c}{Nucleus}&
 \multicolumn{1}{|c}{DD-ME2}&
 \multicolumn{1}{|c}{NL3$^*$}&
  \multicolumn{1}{|c}{FSU}&
   \multicolumn{1}{|c}{DD-PCX}&
   \multicolumn{1}{|c}{TMA}&
   \multicolumn{1}{|c}{FRDM}&
   \multicolumn{1}{|c}{SLy4}&
   \multicolumn{1}{|c|}{HFB}\\
  % \multicolumn{1}{|c|}{Expt.}\\
   \hline
 $^{78}${Zr}&  0.00 &0.00  &0.45  &0.00   &0.49 &0.42 &    &0.17      \\
 $^{80}${Zr}&  0.00 &0.00  &0.46  &0.00   &0.49 &0.43 &0.00&-0.23     \\
 $^{82}${Zr}&  0.00 &0.00  &0.01  &0.00    &0.59 &0.44 &0.00&-0.23    \\
 $^{84}${Zr}&  0.00 &0.00  &0.00  &0.00    &-0.21&-0.24&    &0.13     \\
 $^{86}${Zr}&  0.00 &0.01  &0.00  &0.00    &-0.17&0.01 &    &-0.10    \\
 $^{88}${Zr}&  0.00 &0.00  &0.00  &0.00    &0.00 &-0.01&    &-0.09    \\
 $^{90}${Zr}&  0.00 &0.00  &0.00  &0.00    &-0.02&0.00 &    &0.09     \\
 $^{92}${Zr}&  0.02 &0.00  &0.00  &0.00    &-0.14&0.00 &    &-0.11    \\
 $^{94}${Zr}&  0.20 &0.20  &-0.01 &0.10    &-0.22&-0.16&0.00&-0.15    \\
 $^{96}${Zr}&  0.30 &0.26  &0.17  &0.20    &0.27 &0.24 &-0.15&-0.15   \\
 $^{98}${Zr}&  0.24  &0.51 &0.21  &0.40      & 0.34 &0.34 &-0.20&-0.22\\
 $^{100}${Zr}& 0.30 &0.49  &0.21  &0.50    &0.41 &0.36 &0.42&-0.23    \\
 $^{102}${Zr}& -0.20&0.44  &0.21  &0.40    &0.42 &0.38 &0.43&-0.24    \\
 $^{104}${Zr}& -0.20&0.42  &0.33  &0.40    &0.41 &0.38 &0.43&-0.24    \\
 $^{106}${Zr}& -0.22&0.43  &0.35  &0.40    &0.41 &0.37 &0.42&-0.25    \\
 $^{108}${Zr}& 0.02 &0.42  &0.35  &0.40    &0.41 &0.36 &0.41&-0.24      \\
 $^{110}${Zr}& 0.00 &0.00  &0.33  &0.40    &0.43 &0.36 &0.44&-0.23      \\
 $^{112}${Zr}& -0.02&0.00  &0.00  &0.40    &-0.19&0.36 &0.00&-0.20      \\
 $^{114}${Zr}& 0.00 &0.00  &0.00  &0.10    &-0.17&-0.19&    &-0.20      \\
 $^{116}${Zr}& 0.10 &0.00  &0.00  &0.10    &-0.15&-0.16&    &-0.11      \\
 $^{118}${Zr}& 0.00 &0.00  &0.00  &0.10    &0.00 &-0.15&    &-0.05      \\
 $^{120}${Zr}& 0.02 &0.00  &0.00  &0.00    &0.00 &0.00 &    &0.05           \\
 $^{122}${Zr}& 0.00 &0.00  &0.01  &-0.10   &0.00 &0.00 &    &0.00       \\
 \hline
\hline
\end{tabular}}}
\label{all beta}
\end{table*}

After the multiple cases of shape-coexistence as mentioned in Table \ref{beta}, it is important to compare ground state deformation with other considered parameters and other theories. In Table \ref{all beta}, we have tabulated all these ground state deformation to demonstrate the shape transition especially in Zr isotopes. The results of deformations are found in good match with all the considered parameters.
\begin{figure}[h]
\centering
\includegraphics[width=0.65\textwidth]{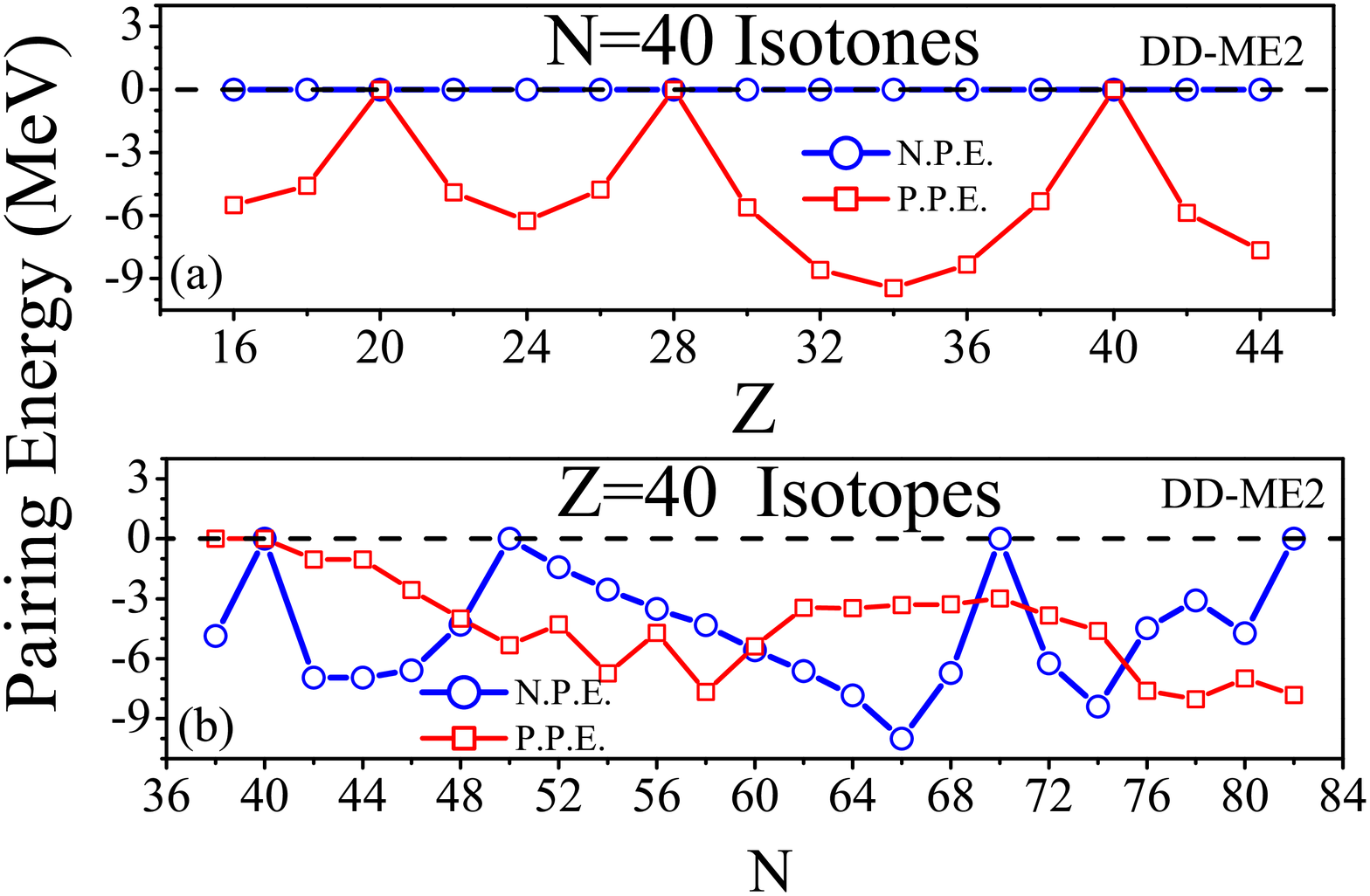}
\caption{(Colour online) Proton and neutron pairing energy (MeV) for N(Z)$=$40 are shown in panels (a) and (b) respectively. Zero value of pairing energy exhibits magicity that results $^{60}$Ca, $^{68}$Ni and $^{80}$Zr as doubly magic nuclei in panel (a). Panel (b) indicates neutron magicity in $^{90,110,122}$Zr as only neutron pairing energy attains zero value for N$=$50, 70 and 82.} \label{fig4}
\end{figure}

\begin{figure}[h]
\centering
\includegraphics[width=0.7\textwidth]{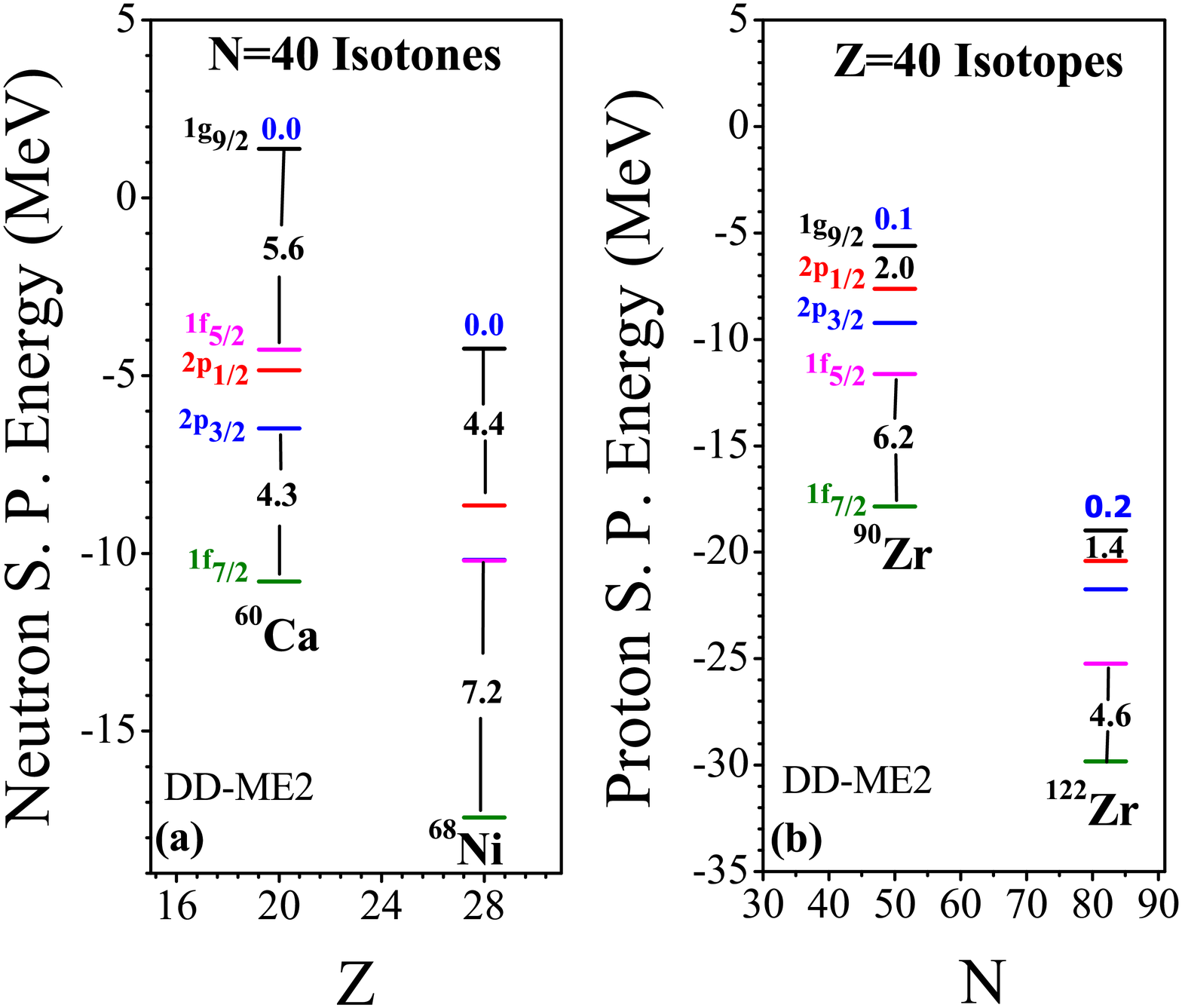}
\caption{(Colour online) Neutron and proton single particle states for N$=$40 and Z$=$40 are shown in panels (a) and (b), respectively. The significant gap between neutron pf-shell and $\nu1g_{9/2}$ confirms magicity at $^{60}$Ca and $^{68}$Ni. In contrast, insignificant gap between proton pf-shell and $\pi1g_{9/2}$ manifests the non-magic character of Zr isotopes.}\label{fig5}
\end{figure}

\subsection{Magicity}
The above detail investigation of shapes reinforce us to examine magicity in these chains of nuclei through the analysis of pairing energy which leads to magic character, if it vanishes ~\cite{Yadav2004,saxena4}, and manifests magicity. In Fig. \ref{fig4}, we have shown the proton and neutron pairing energy contribution for nuclei with N(Z)$=$40 calculated by DD-ME2 parameter. For N$=$40 isotones, neutron pairing energy is found zero for all the nuclei affirming their spherical shapes as found above. However, proton pairing energy reaches to zero along with neutron pairing enery if there is a doubly magic nucleus. This situation indeed arrives for the case of $^{60}$Ca, $^{68}$Ni and $^{80}$Zr which is another confirmation of double magicity. In Fig. \ref{fig4}(b), except $^{80}$Zr, none of the isotopes of Zr is found doubly magic as proton pairing energy does not vanish for any of the isotopes of Zr. However, neutron pairing energy attains zero value for N$=$40, 50, 70 and 82 confirming sphericity of $^{80,90,110,122}$Zr. Importantly, these all nuclei are found with dominant spherical shapes as discussed above in the description of Figs. \ref{fig2} and \ref{fig3}. \par

To get more insight about magicity, we choose the above mentioned magic nuclei to describe magic nature on the basis of single particle energies. For the case of N$=$40 isotones, we have shown neutron single particle levels of $^{60}$Ca and $^{68}$Ni in Fig. \ref{fig5}(a). A large gap between $\nu1f_{7/2}$ and $\nu1f_{5/2}$ can be seen for all these considered nuclei indicating a strong shell closure at N$=$28. The gap which is responsible for shell closure at N$=$40 is the gap between neutorn pf-shell and $\nu1g_{9/2}$ state which is indeed evident from the figure for all these considered nuclei. This gap varies form 4.4 MeV to 5.7 MeV which is quite sizable as that of the gap found for conventional shell closure at N$=$28, therefore, $\nu1g_{9/2}$ state remains vacant as can be seen from the occupancy mentioned (in blue number). This description of single particle levels serves as an accompaniment to strong magicity of $^{60}$Ca and $^{68}$Ni. In a similar manner, proton single particle levels are plotted for the case of Zr isotopes by selecting nuclei $^{90,122}$Zr in Fig. \ref{fig5}(b). As expected, the gap between $\pi1f_{7/2}$ and $\pi1f_{5/2}$ leads to shell closure at Z$=$28, however, the gap within proton pf-shell and $\pi1g_{9/2}$ state is not as large as found for the case of nuclei with N$=$40 and resulting a non-zero occupancy of $\pi1g_{9/2}$. Hence, Zr isotopes are most likely to show non-magic character for proton magicity favouring the outcome obtained from proton pairing energy contribution in Fig. \ref{fig4}(b). Since the gap between single particle states is closely related to pairing correlations therefore the above findings of shell closures should be tested for various other considered models considering the fact that their pairing treatments are different as mentioned above. The gap between pf-shell and $1g_{9/2}$ state is calculated for $^{60}$Ca, $^{68}$Ni, $^{90}$Zr, and $^{122}$Zr nuclei using DD-ME2, NL3*, FSU-Gold, and DD-PCX parameters and has been listed in Table \ref{spe}. The values of gaps from all the considered parameters are found almost similar which establishes the magic character of $^{60}$Ca and $^{68}$Ni along with non-magicity of $^{90}$Zr and $^{122}$Zr. \par

\begin{table*}[!htbp]
\caption{Energy gap between pf-shell and $1g_{9/2}$ state obtained in the RMF calculations using various force parameters.}
\centering
\resizebox{0.75\textwidth}{!}{%
\begin{tabular}{|c|c|c|c|c|}
\hline
\multicolumn{1}{|c|}{Parameters} &
\multicolumn{4}{|c|}{Energy gap between single particle states (MeV)}\\
\cline{2-5}
\multicolumn{1}{|c|}{}  &
\multicolumn{2}{|c|}{Neutron S. P. States} &
\multicolumn{2}{|c|}{Proton S. P. States}\\
\cline{2-5}
\multicolumn{1}{|c|}{}  &
\multicolumn{1}{|c|}{$^{60}$Ca} &
\multicolumn{1}{|c|}{$^{68}$Ni} &
\multicolumn{1}{|c|}{$^{90}$Zr} &
\multicolumn{1}{|c|}{$^{122}$Zr}\\
\hline
DD-ME2   & 5.7&	4.4&	 2.0&	1.4  \\
NL3*     & 5.4&	4.1&	 2.0&	1.8   \\
FSU-Gold & 5.9&	4.2&	 1.9&	1.3   \\
DD-PCX   & 5.9&	4.6&	 2.2&	1.4   \\
\hline
\end{tabular}}
\label{spe}
\end{table*}

\begin{figure}[h]
\centering
\includegraphics[width=0.6\textwidth]{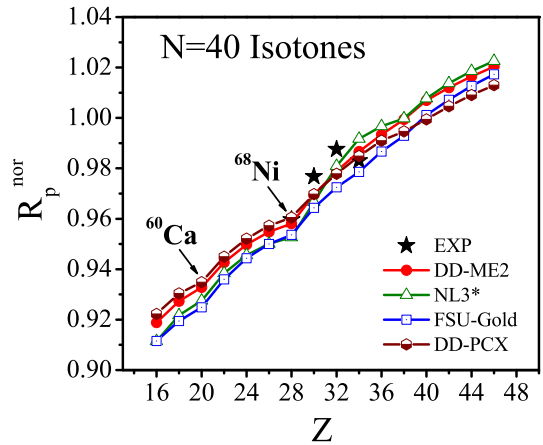}
\caption{(Colour online) Normalized radii (R$_{p}^{nor}$ = R$_{p}$/R$_{p}^{Co}$) for N$=$40 isotones calculated by DD-ME2,  NL3*,  FSU-Gold, and DD-PCX,  parameters. Experimental data are taken from \cite{angeli}. For $^{68}$Ni the experimental data is taken from very recent measurement \cite{kaufmann2020}. The kink observed for $^{60}$Ca and $^{68}$Ni is another demonstration of double magicity in these nuclei from all the considered parameters.} \label{fig6}
\end{figure}

Further, we have also plotted proton rms radii for N$=$40 isotones calculated by DD-ME2, NL3*, FSU-Gold, and DD-PCX parameters. To eliminate the smooth mass number dependence of the proton rms radii R$_{p}$, radii are normalized using the formula reported by Collard \textit{et al.}\cite{collard}.
\begin{equation}
R_{p}^{Co} = \sqrt{3/5}\,\, (1.15 + 1.80A^{-2/3} - 1.20A^{-4/3})\,\,A^{1/3}
\end{equation}
The proton normalized raddi (R$_{p}^{nor}$ = R$_{p}$/R$_{p}^{Co})$ is plotted in the Fig. \ref{fig6}. This plot show a kink (change in the slope) particularly for shell closure \cite{angeli,angeli2} in accordance with the work of Angeli \textit{et al.} and indeed evident in Fig. \ref{fig6} for the case of $^{60}$Ca and $^{68}$Ni. This kind of kink observed for $^{60}$Ca and $^{68}$Ni is another demonstration of double magicity. Moreover, it is gratifying to note that our theoretical results of radius of $^{68}$Ni calculated by DD-ME2 and DD-PCX parameters are in excellent agreement with the very recent measurement by Kaufmann \textit{et al.} \cite{kaufmann2020}. Therefore, from here one can reach to the conclusion that N$=$40 isotones own plenty of grounds for magicity. Various ground state properties obtained with DD-ME2, NL3*, FSU-Gold, and DD-PCX parameters are mentioned in Tables \ref{n=40} and \ref{z=40} with available experimental data \cite{nndc,angeli,kaufmann2020} for N$=$40 isotones and Z$=$40 isotopes, respectively.
\renewcommand{\baselinestretch}{1.1}

\begin{sidewaystable}[bp]%[tbh]
\vspace{400pt}
\caption{Binding energy, charge radius ($R_{c}$), proton radius ($R_{p}$), neutron radius ($R_{n}$), and matter radius ($R_{m}$) for nuclei with N$=$40 are mentioned and compared with available experimental data which are taken from Refs.$~$ \cite{nndc,angeli,kaufmann2020}.}
\centering
\Huge
\resizebox{1.0\textwidth}{!}{%
{\begin{tabular}{|c|c|c|c|c|c|c|c|c|c|c|c|c|c|c|c|c|c|c|c|c|c|c|c|}
 \hline
 \multicolumn{1}{|c}{Nuclei}&
 \multicolumn{5}{|c}{Binding Energy (MeV)}&
 \multicolumn{5}{|c}{$R_{c}$ (fm) }&
 \multicolumn{5}{|c}{$R_{p}$ (fm)}&
 \multicolumn{4}{|c}{$R_{n}$ (fm) }&
 \multicolumn{4}{|c|}{$R_{m}$ (fm) }\\
\cline{2-24}
 \multicolumn{1}{|c}{}&
 \multicolumn{1}{|c}{DDME2}&
  \multicolumn{1}{|c}{NL3$^*$}&
 \multicolumn{1}{|c}{FSU}&
 \multicolumn{1}{|c}{DDPCX}&
 \multicolumn{1}{|c}{Exp.}&
 \multicolumn{1}{|c}{DDME2}&
 \multicolumn{1}{|c}{NL3$^*$}&
   \multicolumn{1}{|c}{FSU}&
    \multicolumn{1}{|c}{DDPCX}&
    \multicolumn{1}{|c}{Exp.}&
    \multicolumn{1}{|c}{DDME2}& \multicolumn{1}{|c}{NL3$^*$}&
       \multicolumn{1}{|c}{FSU}&
        \multicolumn{1}{|c}{DDPCX}&
       \multicolumn{1}{|c}{Exp.}&
          \multicolumn{1}{|c}{DDME2}&
           \multicolumn{1}{|c}{NL3$^*$}&
       \multicolumn{1}{|c}{FSU}&
       \multicolumn{1}{|c}{DDPCX}&
  \multicolumn{1}{|c}{DDME2}&
           \multicolumn{1}{|c}{NL3$^*$}&
  \multicolumn{1}{|c}{FSU}&
  \multicolumn{1}{|c|}{DDPCX} \\
  \hline
\multicolumn{24}{|c|}{N$=$40 Isotones}\\
\hline
$^{56}${S}  &362.16  &371.00 &359.10&360.04 &       &3.54&3.52&3.52&3.56  &    &3.45&3.42  &3.42&3.46  &    &4.15 &4.24  &4.20&4.08  &3.96&4.03&3.99&3.91\\
$^{58}${Ar} &413.99  &419.91 &412.13&412.56 &       &3.61&3.59&3.58&3.62  &    &3.52&3.50  &3.49&3.53  &    &4.10 &4.20  &4.13&4.04  &3.93&3.99&3.94&3.89\\
$^{60}${Ca} &461.71  &465.75 &459.91&460.83 &       &3.66&3.64&3.63&3.67  &    &3.57&3.55  &3.54&3.58  &    &4.06 &4.15  &4.08&4.02  &3.90&3.96&3.96&3.88\\
$^{62}${Ti} &498.42  &500.51 &494.33&498.49 &495.69 &3.73&3.71&3.70&3.74  &    &3.64&3.62  &3.62&3.65  &    &4.03 &4.12  &4.05&4.00  &3.90&3.95&3.90&3.88\\
$^{64}${Cr} &532.27  &532.86 &527.66&532.79 &531.26 &3.79&3.77&3.77&3.80  &    &3.70&3.68  &3.68&3.71  &    &4.01 &4.09  &4.03&3.99  &3.90&3.94&3.90&3.88\\
$^{66}${Fe} &563.40  &562.83 &559.23&564.00 &562.43 &3.84&3.82&3.82&3.85  &    &3.75&3.73  &3.73&3.76  &    &4.00 &4.06  &4.01&3.98  &3.90&3.94&3.94&3.89\\
$^{68}${Ni} &591.87  &590.53 &589.13&592.22 &590.41 &3.88&3.86&3.86&3.89  &3.89&3.80&3.78  &3.78&3.81  &3.81&3.99 &4.04  &3.99&3.97  &3.91&3.93&3.90&3.90\\
$^{70}${Zn} &609.64  &608.57 &605.95&610.44 &611.09 &3.96&3.94&3.93&3.96  &3.98&3.87&3.86  &3.85&3.88  &3.90&4.00 &4.05  &4.00&3.98  &3.95&3.97&3.94&3.94\\
$^{72}${Ge} &625.25  &624.35 &621.67&626.05 &628.69 &4.02&4.03&4.01&4.02  &4.06&3.94&3.95  &3.92&3.94  &3.98&4.02 &4.08  &4.02&4.00  &3.99&4.02&3.97&3.97\\
$^{74}${Se} &638.74  &638.56 &635.89&639.29 &642.89 &4.08&4.10&4.05&4.08  &4.07&4.00&4.02  &3.97&4.00  &3.99&4.04 &4.10  &4.04&4.02  &4.02&4.07&4.00&4.01\\
$^{76}${Kr} &650.13  &649.33 &647.53&650.25 &654.27 &4.14&3.92&4.11&4.13  &4.20&4.06&3.84  &4.04&4.05  &4.12&4.06 &4.54  &4.04&4.04  &4.06&4.33&4.04&4.05\\
$^{78}${Sr} &659.28  &658.99 &658.07&658.68 &663.01 &4.19&4.28&4.17&4.18  &4.26&4.12&4.21  &4.09&4.10  &4.18&4.07 &4.19  &4.05&4.05  &4.09&4.20&4.07&4.08\\
$^{80}${Zr} &665.72  &665.33 &664.75&664.40 &668.80 &4.25&4.26&4.23&4.22  &    &4.18&4.18  &4.15&4.15  &    &4.18 &4.10  &4.07&4.07  &4.14&4.14&4.11&4.11\\
$^{82}${Mo} &666.77  &667.51 &663.78&666.09 &669.37 &4.30&4.31&4.28&4.27  &    &4.23&4.24  &4.21&4.20  &    &4.10 &4.11  &4.08&4.09  &4.17&4.18&4.15&4.14\\
$^{84}${Ru} &665.75  &667.25 &662.20&665.41 &       &4.35&4.36&4.34&4.32  &    &4.28&4.29  &4.26&4.25  &    &4.12 &4.12  &4.10&4.10  &4.20&4.21&4.18&4.18\\
$^{86}${Pd} &662.93  &665.02 &659.44&662.82 &       &4.40&4.41&4.38&4.37  &    &4.32&4.33  &4.31&4.29  &    &4.13 &4.13  &4.11&4.12  &4.23&4.24&4.22&4.21\\
  \hline
 \end{tabular}}}
\label{n=40}
\end{sidewaystable}

\begin{sidewaystable}[bp]%[tbh]
\vspace{300pt}
%\begin{table}
\caption{Binding energy, charge radius ($R_{c}$), proton radius ($R_{p}$), neutron radius ($R_{n}$), and matter radius ($R_{m}$) for Zr isotopes are mentioned and compared with available experimental data which are taken from Refs.$~$ \cite{nndc,angeli,kaufmann2020}.}
\centering
\Huge
\resizebox{1.0\textwidth}{!}{%
{\begin{tabular}{|c|c|c|c|c|c|c|c|c|c|c|c|c|c|c|c|c|c|c|c|c|c|c|c|}
 \hline
 \multicolumn{1}{|c}{Nuclei}&
 \multicolumn{5}{|c}{Binding Energy (MeV)}&
 \multicolumn{5}{|c}{$R_{c}$ (fm) }&
 \multicolumn{5}{|c}{$R_{p}$ (fm)}&
 \multicolumn{4}{|c}{$R_{n}$ (fm) }&
 \multicolumn{4}{|c|}{$R_{m}$ (fm) }\\
\cline{2-24}
 \multicolumn{1}{|c}{}&
 \multicolumn{1}{|c}{DDME2}&
\multicolumn{1}{|c}{NL3$^*$}&
 \multicolumn{1}{|c}{FSU}&
  \multicolumn{1}{|c}{DDPCX}&
 \multicolumn{1}{|c}{Exp.}&
 \multicolumn{1}{|c}{DDME2}&
  \multicolumn{1}{|c}{NL3$^*$}&
   \multicolumn{1}{|c}{FSU}&
    \multicolumn{1}{|c}{DDPCX}&
    \multicolumn{1}{|c}{Exp.}&
    \multicolumn{1}{|c}{DDME2}&
    \multicolumn{1}{|c}{NL3$^*$}&
       \multicolumn{1}{|c}{FSU}&
        \multicolumn{1}{|c}{DDPCX}&
       \multicolumn{1}{|c}{Exp.}&
          \multicolumn{1}{|c}{DDME2}&
           \multicolumn{1}{|c}{NL3$^*$}&
       \multicolumn{1}{|c}{FSU}&
       \multicolumn{1}{|c}{DDPCX}&
  \multicolumn{1}{|c}{DDME2}&
           \multicolumn{1}{|c}{NL3$^*$}&
  \multicolumn{1}{|c}{FSU}&
  \multicolumn{1}{|c|}{DDPCX}\\
  \hline
\multicolumn{24}{|c|}{Z$=$40(Zr) Isotopes}\\
   \hline
  $^{78}${Zr} &635.35 &635.63&635.55&634.55&639.13 &4.25  &4.25 &4.22&4.21  &    &4.17 &4.18 &4.15&4.14 &     &4.05 &4.05 &4.02&4.03 &4.11 &4.12&4.09&4.21\\
  $^{80}${Zr} &665.72 &665.33&664.75&664.40&668.80 &4.25  &4.26 &4.23&4.22  &    &4.18 &4.18 &4.15&4.15 &     &4.18 &4.10 &4.07&4.07 &4.14 &4.14&4.11&4.22\\
  $^{82}${Zr} &690.86 &691.38&693.12&690.66&694.19 &4.26  &4.26 &4.24&4.23  &    &4.18 &4.18 &4.16&4.16 &     &4.17 &4.15 &4.12&4.11 &4.18 &4.17&4.14&4.23\\
  $^{84}${Zr} &714.98 &715.87&717.32&715.51&718.12 &4.26  &4.26 &4.24&4.25  &    &4.18 &4.18 &4.16&4.17 &     &4.17 &4.19 &4.16&4.16 &4.18 &4.19&4.16&4.25\\
  $^{86}${Zr} &738.39 &739.29&740.64&739.23&740.81 &4.26  &4.26 &4.25&4.25  &    &4.18 &4.18 &4.17&4.18 &     &4.21 &4.23 &4.20&4.20 &4.20 &4.21&4.19&4.25\\
  $^{88}${Zr} &761.19 &761.85&763.14&762.25&762.61 &4.26  &4.26 &4.25&4.25  &4.28&4.19 &4.18 &4.18&4.18 &4.20 &4.24 &4.26 &4.23&4.22 &4.22 &4.23&4.21&4.25\\
  $^{90}${Zr} &783.45 &783.50&783.80&784.43&783.90 &4.27  &4.26 &4.26&4.26  &4.27&4.19 &4.18 &4.18&4.19 &4.19 &4.27 &4.29 &4.27&4.26 &4.23 &4.25&4.23&4.26\\
  $^{92}${Zr} &796.39 &796.63&799.67&797.78&799.73 &4.29  &4.28 &4.27&4.28  &4.31&4.21 &4.20 &4.20&4.20 &4.23 &4.33 &4.37 &4.33&4.31 &4.28 &4.30&4.28&4.28\\
  $^{94}${Zr} &810.31 &810.03&812.44&810.73&814.68 &4.34  &4.33 &4.29&4.31  &4.33&4.27 &4.26 &4.21&4.23 &4.26 &4.40 &4.44 &4.40&4.37 &4.34 &4.37&4.32&4.31\\
  $^{96}${Zr} &824.01 &823.51&826.03&823.24&828.99 &4.40  &4.37 &4.37&4.36  &4.35&4.33 &4.30 &4.30&4.29 &4.28 &4.47 &4.51 &4.46&4.43 &4.41 &4.42&4.39&4.36\\
  $^{98}${Zr} &835.46 &836.25&838.85&836.12&840.98 &4.38  &4.50 &4.33&4.47  &4.40&4.31 &4.43 &4.26&4.39 &4.33 &4.50 &4.67 &4.50&4.54 &4.42 &4.57&4.41&4.47\\
  $^{100}${Zr}&847.02 &848.29&850.15&848.81&852.21 &4.45  &4.51 &4.41&4.54  &4.49&4.38 &4.44 &4.34&4.46 &4.42 &4.58 &4.70 &4.55&4.63 &4.50 &4.60&4.47&4.54\\
  $^{102}${Zr}&857.71 &859.28&860.41&859.73&863.57 &4.43  &4.50 &4.38&4.50  &4.53&4.35 &4.43 &4.30&4.43 &4.46 &4.61 &4.73 &4.58&4.62 &4.51 &4.61&4.48&4.50\\
  $^{104}${Zr}&868.06 &869.39&871.30&869.84&873.85 &4.45  &4.51 &4.40&4.52  &    &4.37 &4.44 &4.33&4.45 &     &4.65 &4.77 &4.62&4.66 &4.54 &4.64&4.51&4.52\\
  $^{106}${Zr}&876.35 &878.82&880.62&879.25&882.77 &4.43  &4.54 &4.41&4.55  &    &4.36 &4.47 &4.34&4.48 &     &4.68 &4.82 &4.68&4.70 &4.56 &4.69&4.55&4.55\\
  $^{108}${Zr}&886.13 &887.55&888.94&887.86&891.76 &4.45  &4.55 &4.43&4.57  &    &4.38 &4.48 &4.36&4.50 &     &4.72 &4.86 &4.73&4.74 &4.60 &4.72&4.60&4.57\\
  $^{110}${Zr}&895.22 &896.64&896.07&894.78&899.47 &4.47  &4.45 &4.45&4.59  &    &4.40 &4.38 &4.37&4.52 &     &4.77 &4.86 &4.78&4.78 &4.64 &4.69&4.64&4.59\\
  $^{112}${Zr}&901.41 &904.45&905.46&900.45&906.53 &4.48  &4.46 &4.47&4.60  &    &4.41 &4.39 &4.39&4.53 &     &4.80 &4.90 &4.81&4.83 &4.66 &4.73&4.67&4.60\\
  $^{114}${Zr}&907.06 &911.03&911.59&906.01&       &4.50  &4.48 &4.48&4.51  &    &4.43 &4.41 &4.41&4.44 &     &4.83 &4.94 &4.85&4.77 &4.69 &4.76&4.70&4.51\\
  $^{116}${Zr}&912.45 &916.95&917.04&912.37&       &4.52  &4.50 &4.50&4.52  &    &4.45 &4.42 &4.43&4.45 &     &4.86 &4.97 &4.88&4.80 &4.72 &4.79&4.73&4.52\\
  $^{118}${Zr}&917.67 &922.48&922.02&918.04&       &4.54  &4.52 &4.52&4.54  &    &4.47 &4.44 &4.44&4.47 &     &4.88 &5.01 &4.91&4.83 &4.75 &4.82&4.76&4.54\\
  $^{120}${Zr}&922.35 &927.39&926.50&920.07&       &4.54  &4.53 &4.53&4.55  &    &4.47 &4.46 &4.46&4.48 &     &4.91 &5.04 &4.94&4.84 &4.77 &4.85&4.78&4.55\\
  $^{122}${Zr}&927.02 &931.79&930.22&925.87&       &4.56  &4.55 &4.55&4.58  &    &4.49 &4.48 &4.48&4.51 &     &4.93 &5.07 &4.96&4.89 &4.79 &4.88&4.81&4.58\\
  $^{124}${Zr}&926.62 &932.92&932.82&927.49&       &4.57  &4.56 &4.56&4.58  &    &4.50 &4.49 &4.49&4.51 &     &4.99 &5.14 &5.05&4.92 &4.84 &4.94&4.88&4.58\\
  $^{126}${Zr}&926.15 &933.85&934.23&925.60&       &4.59  &4.57 &4.57&4.60  &    &4.52 &4.50 &4.50&4.53 &     &5.05 &5.20 &5.13&4.98 &4.89 &4.99&4.94&4.60\\
  $^{128}${Zr}&925.60 &934.68&934.77&923.92&       &4.61  &4.58 &4.58&4.61  &    &4.54 &4.51 &4.51&4.54 &     &5.11 &5.27 &5.21&5.03 &4.94 &5.04&5.00&4.61\\
  $^{130}${Zr}&924.76 &935.92&934.82&922.92&       &4.62  &4.63 &4.59&4.78  &    &4.55 &4.56 &4.52&4.71 &     &5.16 &5.34 &5.27&5.19 &4.98 &5.12&5.05&4.78\\
  $^{132}${Zr}&923.68 &936.98&934.54&922.67&       &4.64  &4.65 &4.60&4.79  &    &4.57 &4.58 &4.52&4.73 &     &5.21 &5.40 &5.34&5.23 &5.02 &5.16&5.11&4.79\\
  $^{134}${Zr}&922.38 &937.83&934.06&921.83&       &4.64  &4.67 &4.61&4.81  &    &4.57 &4.60 &4.54&4.74 &     &5.25 &5.45 &5.40&5.27 &5.06 &5.21&5.16&4.81\\
  $^{136}${Zr}&920.94 &938.37&933.35&920.53&       &4.66  &4.69 &4.61&4.83  &    &4.59 &4.62 &4.54&4.76 &     &5.30 &5.50 &5.47&5.30 &5.10 &5.25&5.21&4.83\\
  $^{138}${Zr}&919.29 &938.64&932.44&918.81&       &4.68  &4.70 &4.63&4.85  &    &4.61 &4.64 &4.56&4.78 &     &5.35 &5.54 &5.53&5.34 &5.14 &5.30&5.27&4.85\\
  $^{140}${Zr}&917.48 &938.64&931.34&916.68&       &4.69  &4.72 &4.64&4.86  &    &4.63 &4.65 &4.57&4.80 &     &5.38 &5.59 &5.59&5.38 &5.18 &5.34&5.32&4.86\\
  $^{142}${Zr}&915.51 &938.33&930.04&914.15&       &4.71  &4.74 &4.65&4.88  &    &4.65 &4.67 &4.58&4.81 &     &5.42 &5.63 &5.65&5.42 &5.22 &5.38&5.37&4.88\\
  $^{144}${Zr}&913.43 &938.43&928.60&911.22&       &4.73  &4.69 &4.68&4.86  &    &4.67 &4.62 &4.61&4.80 &     &5.46 &5.67 &5.67&5.38 &5.25 &5.40&5.39&4.86\\
  $^{146}${Zr}&911.23 &938.47&927.05&907.92&       &4.76  &4.71 &4.71&4.91  &    &4.69 &4.64 &4.64&4.85 &     &5.49 &5.71 &5.69&5.50 &5.29 &5.44&5.42&4.91\\
  $^{148}${Zr}&906.91 &938.39&925.18&904.26&       &4.79  &4.73 &4.74&4.93  &    &4.72 &4.67 &4.67&4.86 &     &5.54 &5.75 &5.69&5.54 &5.33 &5.48&5.43&4.93\\
  $^{150}${Zr}&906.36 &938.16&923.20&900.29&       &4.80  &4.76 &4.76&4.94  &    &4.73 &4.69 &4.69&4.88 &     &5.56 &5.78 &5.72&5.58 &5.35 &5.51&5.47&4.94\\
  $^{152}${Zr}&903.43 &937.65&920.98&896.01&       &4.82  &4.78 &4.79&4.96  &    &4.75 &4.71 &4.72&4.90 &     &5.58 &5.81 &5.74&5.63 &5.38 &5.54&5.49&4.96\\
  \hline
\hline
 \end{tabular}}}
\label{z=40}
\end{sidewaystable}
% \end{table}
\renewcommand{\baselinestretch}{1.0}

\subsection{Bubble Structure}

\begin{figure}[ht]
\centering
\includegraphics[width=0.75\textwidth]{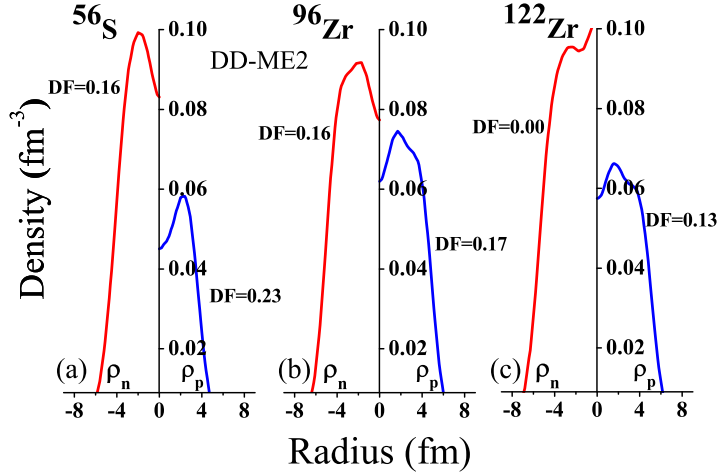}
\caption{(Colour online) Proton and neutron densities of $^{56}_{16}$S, $^{96}_{40}$Zr and $^{122}_{40}$Zr from N(Z)$=$40 isotones(isotopes). The densities are found with significant central depletion in their concern chain of isotones/isotopes. In term of quantity this depletion is also expressed by depletion fraction (DF) ($(\rho_{max}$-$\rho_{c})/\rho_{max}$)~\cite{grasso}, which is also mentioned in the figure.} \label{fig7}
\end{figure}

In addition to the ground state properties and related phenomenon of shape-coexistence and magicity discussed above, there is one more phenomenon which has raised substantial attention  recently after the first experimental proof of central depletion in $^{34}$Si~\cite{nature}. It is the "bubble structure" named to the depletion of central  density of a nuclei. Bubble structure has been discussed in many of the theoretical works~\cite{duguet,todd,grasso,khan,wang,wang1,grasso1,yao1,li ,schuetrumpf,wu} along with our recent work~\cite{saxena,saxena1,saxenajpg,kumawat2020} from which it has been established that $^{34}$Si is a best candidate showing bubble or central depletion in its charge density distribution due to unoccupancy in $\pi2s_{1/2}$. Encouraging with these recent studies, we have analyzed density distribution at center for all the isotones of N$=$40 and isotopes of Z$=$40, which may be analogously influenced by 3s$_{1/2}$ state. From these chains of nuclei, it is found that many of the nuclei indeed possess the central depletion in their proton or neutron or both the densities. To visualize this phenomenon, in Fig. \ref{fig7}, we have shown proton and neutron densities together of some selected nuclei which are found with significant central depletion in their concern chain of isotones/isotopes. The diminution in the density at center can be express as a number defined as depletion fraction (DF) ($(\rho_{max}$-$\rho_{c})/\rho_{max}$, where $\rho_{max}$ is maximum density and $\rho_{c}$ is the density at center)~\cite{grasso}, which is also mentioned in the Fig. \ref{fig7}. From N$=$40 isotones, we have selected $^{56}_{16}$S and from Zr isotopes the isotopes considered are $^{96}_{40}$Zr and $^{122}_{40}$Zr. Bubble phenomenon is clearly observed in both neutron and proton densities with a reasonable value of DF as mentioned in the plots. However, for the case of $^{122}$Zr, neutron DF is found zero which is oblivious as no depletion in neutron density is observed. \par

\begin{figure}[h]
\centering
\includegraphics[width=0.6\textwidth]{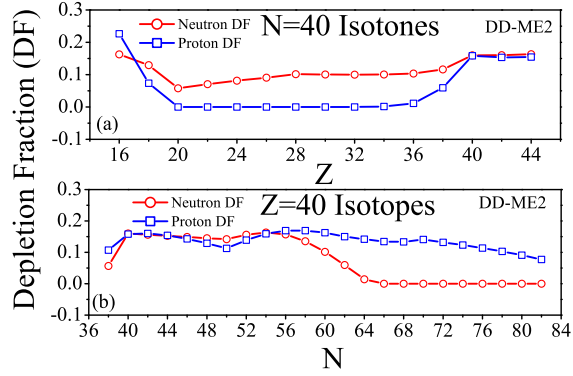}
\caption{(Colour online) Proton and neutron depletion fraction (DF) ($(\rho_{max}$-$\rho_{c})/\rho_{max}$)~\cite{grasso} for N(Z)$=$40 isotones(isotopes). From panel (a) neutron DF is found consistent in N$=$40 isotones, though with a low value, whereas proton DF is dependent on proton number Z. Panel (b) depicts substantial bubble structure in the proton density of Zr isotopes.} \label{fig8}
\end{figure}

To perform a systematic study of depletion in proton and neutron densities, in Fig. \ref{fig8}, we have displayed proton and neutron depletion fraction (DF) for full chain of isotones of N$=$40 and isotopes of Zr. From Fig. \ref{fig8}(a), neutron DF is found consistent in N$=$40 isotones, though with a low value, whereas proton DF is ascertained with the dependency on proton number Z. Proton depletion is found maximum for $^{56}_{16}$S which is due to vacant $\pi2s_{1/2}$ state and reaches to zero for Z$\geq$20 as $\pi2s_{1/2}$ state occupies completely. The proton DF remains zero upto Z$<$40 and then increases after Z$\geq$40 which may be anticipated with the inclusion of vacant $\pi3s_{1/2}$ state while filling of $\pi1g_{9/2}$ state. For the case of Zr isotopes, this bubble phenomenon of proton density is found in a more general manner as can be seen from Fig. \ref{fig8}(b), which fortifies the role of unoccupied $\pi3s_{1/2}$ state in this region of periodic chart. Towards neutron side this role of $3s_{1/2}$ state determining central depletion can be easily visualized for which we have plotted occupancies of $\nu1g_{9/2}$, $\nu3s_{1/2}$ and $\nu1h_{11/2}$ single particle states for all the isotopes of Zr.\par

\begin{figure}[h]
\centering
\includegraphics[width=0.55\textwidth]{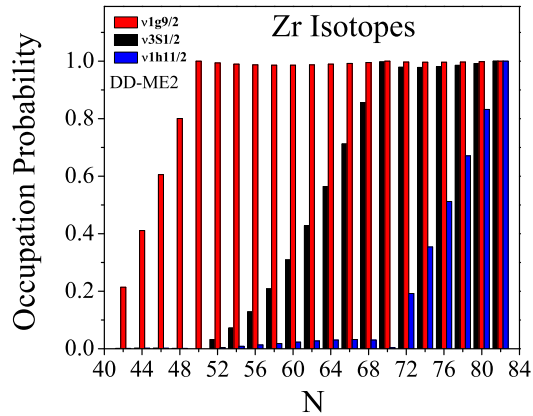}
\caption{(Colour online) Occupation probability of $\nu1g_{9/2}$, $\nu3s_{1/2}$ and $\nu1h_{11/2}$ of Zr isotopes. The figures evidently describes the role of $\nu3s_{1/2}$ state in the bubble structure of Zr isotopes.} \label{fig9}
\end{figure}

From Fig. \ref{fig9}, one can see that as neutron number increases for N$>$40, $\nu1g_{9/2}$ state starts to accommodate with the neutrons and fills completely at N$=$50, however, upto this neutron number the $\nu3s_{1/2}$ state remains unoccupied. After N$>$50, $\nu3s_{1/2}$ state starts to fill gradually along with few other single particle states viz. $\nu1g_{7/2}$, $\nu2d_{5/2}$ and $\nu2d_{3/2}$. As one reaches to N$\geq$64, $\nu3s_{1/2}$ state has its occupancy $>$ 0.5 leading to zero DF as can be seen in Fig. \ref{fig8}(b). This outcome where DF reaches to zero after 50\% of occupancy of s-state is in agreement with our earlier work for $2s_{1/2}$ state \cite{saxenajpg}. Therefore, after N$\geq$64, DF reaches to zero and subsequently $\nu3s_{1/2}$ and $\nu1h_{11/2}$ single particle states get occupied upto N$=$82, hence, no depletion has been found in the neutron density of $^{122}_{40}$Zr as seen in Fig. \ref{fig7}. Dependency of DF on various parameters of RMF has been discussed in detail already \cite{saxenajpg}, however, a deeper and separate study of bubble phenomenon in this region of periodic chart is required.

\section{Conclusions}
We describe ground state properties of N$=$40 isotones and Z$=$40(Zr) isotopes through investigating separation energies, deformation, single particle energies, pairing energies, radii, proton and neutron density profiles of even-even nuclei. For this systematic study, we employ various relativistic mean-field (RMF) models comprising density dependent meson exchange model (DD-ME2), density dependent point coupling interaction (DD-PCX) and the model with nonlinear self- and mixed-interactions of the mesons viz.  NL3* and FSU-Gold. The obtained results are compared within the RMF models, available experimental data and various other theories. N$=$40 isotones own strong candidature of magicity which declares $^{60}$Ca and $^{68}$Ni as doubly magic nuclei, whereas, most of the Zr isotopes are found deformed in which shape transition and shape co-existence are clearly observed. Among the nuclei from both chains, few nuclei are identified which show both proton and neutron central depletion in density distribution referred as doubly bubble nuclei. Evidentially, $^{56}_{16}$S and $^{122}_{40}$Zr are reported as doubly bubble nuclei.

\section{Acknowledgement}
Authors would like to thank Prof. B. K. Agarwal, SINP, India for his kind guidance and continuous support. Authors are also grateful to M. Kumawat for her help in the preparation of this manuscript. The authors take great pleasure in thanking the referee for his several suggestions and comments which helped to improve the manuscript. The financial support provided by SERB (DST), Govt. of India under CRG/2019/001851 is gratefully acknowledged.


\begin{thebibliography}{000}
\bibitem{tarasov2018} O. B. Tarasov \textit{et al.}, \textit{Phys. Rev. Lett.} \textbf{121}, (2018) 022501.
\bibitem{purnima2018} P. Singh \textit{et al.}, \textit{Phys. Rev. Lett.} \textbf{121}, (2018) 192501.
\bibitem{cortes2020} M. L. Cortés \textit{et al.}, \textit{Phys. Lett. B} \textbf{800}, (2020) 135071.
\bibitem{izzo2018} C. Izzo \textit{et al.}, \textit{Phys. Rev. C} \textbf{97}, (2018) 014309.
\bibitem{flavigny2019} F. Flavigny \textit{et al.}, \textit{Phys. Rev. C} \textbf{99}, (2019) 054332.
\bibitem{bonasera2018} G. Bonasera, M. R. Anders, and S. Shlomo, \textit{Phys. Rev. C} \textbf{98}, (2018) 054316.
\bibitem{martorana2018} N. S. Martorana \textit{et al.}, \textit{Phys. Lett. B} \textbf{782}, (2018) 112.
\bibitem{sun2018}  Xuwei Sun, Jing Chen, and Dinghui Lu. \textit{Phys. Rev. C} \textbf{98}, (2018) 024607.
\bibitem{gavrielov2019} N. Gavrielov, A. Leviatan and F. Iachello, \textit{Phys. Scr.} \textbf{95} (2019) 024001.
\bibitem{wang2015} Z. H. Wang, J. Xiang, W. H. Long and Z. P. Li, \textit{J. Phys. G: Nucl. Part. Phys.} \textbf{42} (2015) 045108.
\bibitem{ramos2019} J. E. García-Ramos and K. Heyde, \textit{Phys. Rev. C} \textbf{100}, (2019) 044315.
\bibitem{miyahara2018} S. Miyahara and H. Nakada, \textit{Phys. Rev. C} \textbf{98}, (2018) 064318.
\bibitem{adrich2008} P. Adrich \textit{et al.}, \textit{Phys. Rev. C} \textbf{77}, (2008) 054306.
\bibitem{ljungvall2010}J. Ljungvall \textit{et al.}, \textit{Phys. Rev. C} \textbf{81}, (2010) 061301(R).
\bibitem{pauwels2008} D. Pauwels \textit{et al.}, \textit{Phys. Rev. C} \textbf{78}, (2008) 041307(R).
\bibitem{lenzi2010} S. M. Lenzi, F. Nowacki, A. Poves, and K. Sieja, \textit{Phys. Rev. C} \textbf{82}, (2010) 054301.
\bibitem{gaude2009} L. Gaudefroy \textit{et al.}, \textit{Phys. Rev. C} \textbf{80}, (2009) 064313.
\bibitem{naimi2012} S. Naimi \textit{et al.}, \textit{Phys. Rev. C} \textbf{86}, (2012) 014325.
\bibitem{liu2018} X. Y. Liu \textit{et al.}, \textit{Phys. Lett. B} \textbf{784}, (2018) 392.
\bibitem{wimmer2019} K. Wimmer \textit{et al.}, \textit{Phys. Lett. B} \textbf{792}, (2019) 16.
\bibitem{brown2002}  B. A. Brown, R. R. C. Clement, H. Schatz, A. Volya, and W. A. Richter, \textit{Phys. Rev. C} \textbf{65}, (2002) 045802.
\bibitem{otsuka2020} T. Otsuka, A. Gade, O. Sorlin, T. Suzuki, and Y. Utsuno, \textit{Rev. Mod. Phys.} textbf{92}, (2020) 015002.
\bibitem{Lalazissis05} G. A. Lalazissis, T. Niksic, D. Vretenar, and P. Ring, \textit{Phys. Rev. C} \textbf{71}, (2005) 024312.
\bibitem{ddpcx}E. Yüksel, T. Marketin and N. Paar, \textit{Phys. Rev. C} \textbf{99}, (2019) 034318.
\bibitem{Boguta77} J. Boguta and A. R. Bodmer, \textit{Nucl. Phys. A} \textbf{292}, (1977) 413.
\bibitem{Boguta83} J. Boguta and H. Stoecker, \textit{Phys. Lett. B} \textbf{120}, (1983) 289.
\bibitem{Furnstahl97} R. J. Furnstahl, B. D. Serot, and H. B. Tang, \textit{Nucl. Phys. A} \textbf{615}, (1997) 441.
\bibitem{Lalazissis09} G. A. Lalazissis, S. Karatzikos, R. Fossion, D. Pena Arteaga, A. V. Afanasjev and P. Ring, \textit{Phys. Lett. B} \textbf{671},
(2009) 36.
\bibitem{Todd-Rutel05} B. G. Todd-Rutel and J. Piekarewicz, \textit{Phys. Rev. Lett.} \textbf{95}, (2005) 122501.
\bibitem{nndc} https://www.nndc.bnl.gov/nudat2/
\bibitem{angeli} I. Angeli and K. P. Marinova, \textit{Atomic Data and Nuclear Data Tables} \textbf{99}, (2013) 69.
\bibitem{agbemava2014} S. E. Agbemava, A. V. Afanasjev, D. Ray, and P. Ring, \textit{Phys. Rev. C} \textbf{89}, (2014) 054320.
\bibitem{kucharek1991} H. Kucharek and P. Ring,  \textit{Z. Phys. A} \textbf{339}, (1991) 23.
\bibitem{ring1996} P. Ring, \textit{Prog. Part. Nucl. Phys.} \textbf{37}, (1996) 193.
\bibitem{afanasjev2000} A. V. Afanasjev, P. Ring, and J. Konig, \textit{Nucl. Phys. A} \textbf{676}, (2000) 196.
\bibitem{saxena4}G. Saxena, D. Singh, M. Kaushik, H. L. Yadav, and H. Toki, \textit{Int. Jour. Mod. Phys. E} \textbf{22}, (2013) 1350025.
\bibitem{Yadav2004} H. L. Yadav, M. Kaushik, and H. Toki, \textit{Int. J. Mod. Phys. E} \textbf{13}, (2004) 647.
\bibitem{Dobaczewski1983}J. Dobaczewski, H. Flocard and J. Treiner, \textit{Nuclear Physics A} \textbf{422} (1984) 103.
\bibitem{Bertsch1991}G. F. Bertsch and H. Esbensen, \textit{Annals Phys.} \textbf{209}, (1991) 327.
\bibitem{Dobaczewski1995}J. Dobaczewski, W. Nazarewicz, T. R. Werner, J. F. Berger, C. R. Chinn and J. Decharge, \textit{Phys. Rev. C} \textbf{53}, (1996) 2809.
\bibitem{Gambhir1989}Y. K. Gambhir, P. Ring and A. Thimet, \textit{Annals Phys.} \textbf{198}, (1990) 132.
\bibitem{Singh2013}D. Singh, G. Saxena, M. Kaushik, H. L. Yadav and H. Toki, \textit{Int. J. Mod. Phys. E} \textbf{21},(2012) 9.
\bibitem{Geng2003}L.-S. Geng, H. Toki, S. Sugimoto and J. Meng, \textit{Prog. Theor. Phys.} \textbf{110}, (2003) 921.
\bibitem{reinhard1986} P. -G. Reinhard, M. Rufa, J. Maruhn, W. Greiner, and J. Friedrich, \textit{Z. Phys. A} \textbf{323}, (1986) 13.
\bibitem{suga1994} Y. Sugahara and H. Toki, \textit{Nucl. Phys. A} \textbf{579}, (1994) 557.
\bibitem{TMR} Y. Tian, Z. Y. Ma, and P. Ring, \textit{Phys. Lett. B} \textbf{676}, (2009) 44.
\bibitem{berger1991} J. F. Berger, M. Girod, and D. Gogny, \textit{Comp. Phys. Comm.} \textbf{63}, 365 (1991).
\bibitem{meng1998} J. Meng and P. Ring, \textit{Phys. Rev. Lett.} \textbf{80}, (1998) 460.
\bibitem{akbari} M. Akbari, A. Kardan, \textit{Nucl. Phys. A} \textbf{990} (2019) 109.
\bibitem{nomura}K. Nomura \textit{et al.}, \textit{Phys. Rev. C} \textbf{95}, (2017) 064310.
\bibitem{ayangeakaa} A. D. Ayangeakaa \textit{et al.}, \textit{Phys. Lett. B} \textbf{754}, (2016) 254.
\bibitem{delaroche2010} J.-P. Delaroche, M. Girod, J. Libert, H. Goutte, S. Hilaire, S. Peru, N. Pillet and G.F. Bertsch, \textit{Phys. Rev. C} \textbf{81} (2010) 014303.
\bibitem{Bender09}M. Bender, K. Bennaceur, T. Duguet, P. H. Heenen, T. Lesinski and J. Meyer \textit{Phys. Rev. C} \textbf{80}, (2009) 064302.
\bibitem{Zou2010}W. -H. Zou, Y. Tian, S. -F. Shen, J. -Z. Gu, B. -B. Peng, D. -D.Zhang and Z.-Y. Ma \textit{Chin.Phys.C} \textbf{34}, (2010) 56.
\bibitem{Zheng2014}S. J. Zheng, F. R. Xu, S. F. Shen, H. L. Liu, R. Wyss and Y.P.Yan \textit{Phys.Rev. C} \textbf{90}, (2014) 064309.
\bibitem{Sazo} D. A. Sazonov, E. A. Kolganova, T. M. Shneidman, R. V. Jolos, N. Pietralla, and W. Witt, \textit{Phys. Rev. C} \textbf{99}, (2019) 031304.
\bibitem{witt}W. Witt \textit{et al.}, \textit{Phys. Rev. C} \textbf{98}, (2018) 041302(R).
\bibitem{kremer} C. Kremer \textit{et al.}, \textit{Phys. Rev. Lett.} \textbf{117}, (2016) 172503.
\bibitem{YanXin} L. YanXin, Y. ShaoYing and S. Yang, \textit{Sci. China Phys. Mech. Astron.} \textbf{58}, (2015) 112003.
\bibitem{geng2004} L. S. Geng, H. Toki and J. Meng, \textit{Mod. Phys. Lett. A} \textbf{19}, (2004) 2171-2190.
\bibitem{frdm2012} P. Moller, A. J. Sierk, T. Ichikawa, and H. Sagawa, \textit{Atomic Data and Nuclear Data Tables} \textbf{109–110}, (2016) 1.
\bibitem{goriely} S. Goriely \textit{et al.}, \textit{Phys. Rev. C} \textbf{88}, (2013) 024308.
\bibitem{goriely-xu} Yi Xu, S. Goriely, A. Jorissen, G. Chen, M. Arnould, \textit{Astronomy \& Astrophysics} \textbf{549}, (2013) A106.
\bibitem{bharat2015} B. Kumar, S. K. Singh and S. K. Patra, \textit{Int. J. Mod. Phys. E} \textbf{24}, (2015) 1550017.
\bibitem{kaufmann2020} S. Kaufmann \textit{et al.}, \textit{Phys. Rev. Lett.} \textbf{124}, (2020) 132502.
\bibitem{collard} H. R. Collard, L. R. B. Elton, and R. Hofstadter, \textit{Springer Berlin}  \textbf{2}, (1967).
\bibitem{angeli2} I. Angeli and K. P. Marinova, \textit{J. Phys. G: Nucl. Part. Phys.} \textbf{42}, (2015) 055108.
\bibitem{nature} A. Mutschler \textit{et al.}, \textit{Nature Physics} \textbf{13}, (2017) 152.
\bibitem{duguet} T. Duguet, V. Soma, S. Lecluse, C. Barbieri, and P. Navratil, \textit{Phys. Rev. C} \textbf{95}, (2017) 034319.
\bibitem{todd} B. G. Todd-Rutel, J. Piekarewicz, P. D. Cottle, \textit{Phys. Rev. C} \textbf{69}, (2004) 021301(R).
\bibitem{grasso} M. Grasso, Z. Ma, E. Khan, J. Margueron, Nguyen Van Giai, \textit{Phys. Rev. C}  \textbf{76}, (2007) 044319.
\bibitem{khan} E. Khan, M. Grasso, J. Margueron, N. Van Giai, \textit{Nucl. Phys. A} \textbf{800}, (2008) 37.
\bibitem{wang} Y. Z. Wang, J. Z. Gu, X. Z. Zhang, J. M. Dong, \textit{Chin. Phys. Lett.} \textbf{28}, (2011) 102101.
\bibitem{wang1} Y. Z. Wang, J. Z. Gu, X. Z. Zhang, J. M. Dong, \textit{Phys.Rev. C} \textbf{84}, (2011) 044333.
\bibitem{grasso1} M. Grasso \textit{et al.}, \textit{Phys. Rev. C} \textbf{79}, (2009) 034318.
\bibitem{yao1} Yao, Jiang-Ming \textit{et al.}, \textit{Phys. Rev. C} \textbf{86}, (2012) 014310.
\bibitem{li} J. J. Li, W. H. Long, J. L. Song, and Q. Zhao,  \textit{Phys. Rev. C} \textbf{93}, (2016) 054312.
\bibitem{schuetrumpf} B. Schuetrumpf, W. Nazarewicz, and P. G. Reinhard, \textit{Phys. Rev. C} \textbf{96}, (2017) 024306.
\bibitem{wu} X. Y. Wu, J. M. Yao, and Z. P. Li, \textit{Phys. Rev. C}  \textbf{89}, (2014) 017304.
\bibitem{saxena} G. Saxena, M. Kumawat, M. Kaushik, S. K. Jain, and Mamta Aggarwal, \textit{Phys. Lett. B} \textbf{788}, (2019) 1.
\bibitem{saxena1} G. Saxena, M. Kumawat, B. K. Agrawal and Mamta Aggarwal, \textit{Phys. Lett. B} \textbf{789}, (2019) 323.
\bibitem{saxenajpg} G. Saxena, M Kumawat, B. K. Agrawal and Mamta Aggarwal, \textit{J. Phys. G: Nucl. Part. Phys.} \textbf{46}, (2019) 065105.
\bibitem{kumawat2020}M. Kumawat, G. Saxena, M. Kaushik, S. K. Jain, J. K. Deegwal and Mamta Aggarwal, \textit{Int. Jour. Mod. Phys. E} \textbf{29}, (2020) 2050068.
\end{thebibliography}
\end{document}